\documentclass[10pt,twocolumn]{article}
\usepackage[top=1.9cm, bottom=1.9cm, left=1.7cm, right=1.7cm]{geometry}
\usepackage{times}
\usepackage[hyphens]{url}  %
\usepackage[keeplastbox]{flushend}  %
\usepackage{graphicx}  %
\usepackage{times}  
\usepackage{helvet}
\usepackage{courier} 
\usepackage[hyphens]{url}  
\usepackage{graphicx} 

\usepackage{amsmath}
\usepackage{url}

\usepackage{xcolor}
\usepackage{flushend}
\usepackage{booktabs}
\usepackage{paralist}
\usepackage{enumitem}
\usepackage[small,bf]{caption}
\usepackage{caption}
\usepackage{subcaption}
\usepackage[hang,flushmargin]{footmisc}

\usepackage[bookmarks=true,%
            bookmarksnumbered=true,%
            colorlinks=true,%
            linkcolor=linkcol,%
            citecolor=citecol,%
            urlcolor=urlcol,%
            hypertexnames=true,%
            pdfpagelabels,
	    ]%
      {hyperref}

\usepackage[OT2,OT1]{fontenc}
\newcommand\cyr
{
\renewcommand\rmdefault{wncyr}
\renewcommand\sfdefault{wncyss}
\renewcommand\encodingdefault{OT2}
\normalfont
\selectfont
}
\DeclareTextFontCommand{\textcyr}{\cyr}

\usepackage[compact]{titlesec}
\titlespacing*{\section}{0pt}{*4}{4pt} 
\titlespacing{\subsection}{0pt}{*3}{3pt}

\definecolor{linkcol}{rgb}{0,0,0.7}
\definecolor{citecol}{rgb}{0,0,0.7}
\definecolor{urlcol}{rgb}{0,0,0.7}

\renewenvironment{thebibliography}[1]{
  \begin{oldthebibliography}{#1}
    \setlength{\itemsep}{0.1em}
    \setlength{\parskip}{0.1em}
}
{
  \end{oldthebibliography}
}

\renewcommand{\footnoterule}{%
  \kern -3pt
  \hrule width 1in
  \kern 2pt
}

\usepackage{xspace}

\makeatletter
\def\url@leostyle{%
  \@ifundefined{selectfont}{\def\UrlFont{}}%
  {\def\UrlFont{}}%
}
\makeatother
\usepackage[hyphenbreaks]{breakurl}

\newif\ifwatermark
\watermarkfalse

\ifwatermark
    \usepackage{draftwatermark}
    \SetWatermarkScale{.7}
    \SetWatermarkText{\shortstack{In Submission:\\Do Not Distribute}}
    \SetWatermarkColor[rgb]{1,0.88,0.88}
\fi

\usepackage{etoolbox}
\makeatletter
\patchcmd\@combinedblfloats{\box\@outputbox}{\unvbox\@outputbox}{}{%
   \errmessage{\noexpand\@combinedblfloats could not be patched}%
}%
 \makeatother
\AtBeginShipout{%
  \ifnum\value{page}>1 %
    \typeout{* Additional boxing of page `\thepage'}%
    \setbox\AtBeginShipoutBox=\hbox{\copy\AtBeginShipoutBox}%
  \fi
}

\begin{document}
\title{\LARGE \bf On the Influence of Twitter Trolls during the 2016 US Presidential Election}

\author{\Large \textbf{Nikos Salamanos,\textsuperscript{\rm 1}  Michael J. Jensen,\textsuperscript{\rm 2} Xinlei He,\textsuperscript{\rm 3} Yang Chen,\textsuperscript{\rm 4} Michael Sirivianos\textsuperscript{\rm 5}} \\
	\textsuperscript{\rm 1,5}Cyprus University of Technology \{nik.salaman@cut.ac.cy; michael.sirivianos@cut.ac.cy\}\\
	\textsuperscript{\rm 2}University of Canberra \{Michael.Jensen@canberra.edu.au\}\\
	\textsuperscript{\rm 3,4}Fudan University \{ xlhe17@fudan.edu.cn; chenyang@fudan.edu.cn\ \}
}
\date{}

\maketitle
\begin{abstract}
It is a widely accepted fact that state--sponsored Twitter accounts operated during the 2016 US presidential election spreading millions of tweets with misinformation and inflammatory political content. Whether these social media campaigns of the so--called ``troll'' accounts were able to manipulate public opinion is still in question. Here we aim to quantify the influence of troll accounts and the impact they had on Twitter by analyzing 152.5 million tweets from 9.9 million users, including 822 troll accounts. The data collected during the US election campaign, contain original troll tweets before they were deleted by Twitter. From these data, we constructed a very large interaction graph; a directed graph of 9.3 million nodes and 169.9 million edges. Recently, Twitter released datasets on the misinformation campaigns of 8,275 state--sponsored accounts linked to Russia, Iran and Venezuela as part of the investigation on the foreign interference in the 2016 US election. These data serve as ground--truth identifier of troll users in our dataset. Using graph analysis techniques we qualify the diffusion cascades of web and media context that have been shared by the troll accounts. We present strong evidence that authentic users were the source of the viral cascades. Although the trolls were participating in the viral cascades, they did not have a leading role in them and only four troll accounts were truly influential.    
\end{abstract}

\section{Introduction} 
\label{sec:intro}
The Russian efforts to interfere in and manipulate the outcome of the 2016 US presidential election were unprecedented in terms of the size and scope of the operations. Millions of posts across multiple social media platforms gave rise to hundreds of millions of impressions targeting specific segments of the population in an effort to mobilize, suppress, or shift votes~\cite{Jamieson:2018}. Trolls were particularly focused on the promotion of identity narratives~\cite{Jensen:2018}, though that does not distinguish them from many other actors during the election \cite{Sides:2018}. The Special Counsel's report described this interference as "sweeping and systematic"~(\cite{Mueller:2019}, vol 1, 1). Russia demonstrated an impressive array of tactics for producing significant damage to the integrity of the communication spaces where Americans became informed and discussed their political choices during the election \cite{Mazarr:2019}. 
 
While Russia's efforts continue "unabated"~\cite{Wray:2019}, it is likely they and others will seek to target the American election in 2020 as well as continuing to target elections in Europe and elsewhere. It is important, therefore to identify the operational tactics of social media influence operations if we are to promulgate adequate defenses against them in the future. 

There is a considerable debate as to whether state--sponsored disinformation campaigns that operated on social media were able to affect the outcome of the 2016 US Presidential election. While there is a large body of work which tried to address this question from distinct disciplinary angles \cite{Benkler:2018,Jamieson:2018,Sides:2018} a conclusive result is still missing. There are several obstacles that any empirical study on this subject has dealt with: (i) the lack of complete and unbiased Twitter data -- the Twitter API returns only a small sample of the users' daily activity; (ii) The tweets from deactivated profiles are not available; (iii) The followers and followees lists are not always accessible, hence the social graph is unknown. Moreover, the disinformation strategies that the operators of the state--sponsored accounts had employed is also unclear. A naive approach such as the one that flooded the network with fake--news is not obviously the case. A study of Russian social media activity has found that the majority of the communications are not obviously false~\cite{Schafer:2018}. It is equally possible that the operators had performed advanced manipulation techniques such as first building a reliable social profile aiming to engage a group of followers and then transmitting factually correct, but otherwise deceptive and manipulative claims advancing the political objectives of the disinformation campaign. Hence, text mining and machine learning techniques might not perform well under this scenario.

In this paper we measure the impact of troll activities on the virality of the ambiguous political information that had been shared on Twitter during the 2016 US Presidential election. For that purpose, a very large directed graph has been constructed by the interactions between the users (tweet replies and mentions). The graph consists of 9.3 million nodes and 169 million edges and it has been constructed based on two Twitter datasets: (i) A collection of 152.5 million tweets that was downloaded using the Twitter API during the US presidential election period (from September 21 to November 7, 2016). Hence, we have access to original troll tweets that have yet to be deleted by Twitter. (ii) A collection of original troll tweets which have been released by Twitter itself as part of the investigation on foreign interference in the 2016 US election -- the misinformation campaigns of 8,275 state--sponsored accounts linked to Russia, Iran and Venezuela states. Using graph analysis techniques and classification, we are able to identify the group of users that were most probably the driving force of the viral cascades that have been shared by the troll accounts. 

\textbf{Contribution:} Our primary contributions are as follows:
\begin{itemize}
	\item We construct one of the largest graphs in the literature, which represents the interactions between state--sponsored troll accounts and authentic users in Twitter during the period of 2016 US Presidential election. This is an approximation of the original followers--followees social graph. 
	\item We present strong evidence that the trolls' activity was not the main cause that led to viral cascades of web and media material in Twitter. The experimental results clearly show that the authentic users who had close proximity with the trolls in the graph were the most active and influential part of the population and their activity was the driving force of the viral materials. A possible scenario is that instead of injecting new content, these trolls bandwagon or "resonate" with communities online with which they sought to form relationships. They do so, particularly by targeting opinion leaders in these communities. This is consistent with previous literature on information warfare tactics \cite{Clark:2017}.   
\end{itemize}

\section{Related Work}\label{sec:literature} 

\subsection{Diffusion of disinformation}

In \cite{Vosoughi:2018} the authors investigated the diffusion cascades of true and false rumors distributed on Twitter from 2006 to 2017; approximately 126,000 rumor cascades which have been spread by 3 million people. The rumors had been verified as true or false by six fact--checking organizations. The main funding of this study is that false news diffused faster and more broadly than the truth and also human behavior contributes more to the spread of falsity than the trolls.

\cite{Bovet:2019} examined 171 million tweets collected during five months prior to the 2016 US presidential election. From this collection, they analyzed 30 million tweets shared by 2.3 million users that contained at least one web--URL linking to a news outlet website (outside of Twitter). The 25\% of these news were either fake or biased representing the spreading misinformation on Twitter. Then, in order to investigate the flow of information, the retweet networks are constructed for each category of news. Two users $i$ and $j$ are connected by the directed edge $(j, i)$ if user $i$ retweeted a tweet of user $j$. Hence, the edges represent the direction of information flow.

\cite{Grinberg2019} investigated the extent to which Twitter users had been exposed to fake news during the 2016 US presidential election. The findings suggest that only a small fraction of 1\% of the population was responsible for the diffusion of 80\% of the fake news. Moreover, they proposed policies which had they been adopted by the social media platforms they would have reduced the spread of disinformation.

In~\cite{Zannettou:www2019,Zannettou:WebSci2019} the authors analyzed the characteristics and strategies of 5.5K Russian and Iranian troll accounts in Twitter and Reddit. Moreover, using the \textit{Hawkes Process} they compute an overall statistical measure of influence that quantifies the effect these accounts had on social media platforms, such as Twitter, Reddit, 4chan and Gab. 

\cite{Badawy:2019} examined the Russian disinformation campaigns in the 2016 US election on Twitter. The study is based on 43 million posts shared on Twitter by 5.7 million distinct users (September 16 to November 9, 2016). The study focuses on the characteristics of the \textit{spreaders}, namely the users that had been exposed and shared content produced by Russian trolls. They showed that existing techniques such as the \textit{Botometer} \cite{Davis:2016} are able to effectively identify troll accounts.

An analysis of the role of Russian trolls on Twitter during the 2016 US election is presented in~\cite{Kim:2019}. The \textit{time-sensitive semantic edit distance} (a text distance metric) is proposed for the visualization and qualitative analysis of trolls' strategies such as \textit{left--leaning} and \textit{right--leaning}. 

\subsection{Identifying malicious activity}
A well-known method for identifying troll accounts on Twitter is the \textit{Botometer} (a.k.a. BotOrNot) introduced by \cite{Davis:2016}. The Botometer is a publicly available platform for estimating whether existing Twitter accounts have the characteristics of social bots. Finally, in \cite{Vosoughi:2017} the \textit{Rumor Gauge} is proposed; a method for predicting the veracity of rumors in Twitter during real--world events. It is a system for automatic verification of rumors in real--time, before verification by trusted channels such as governmental organizations is performed.

\section{Datasets} \label{sec:data} 

\subsection{Ground--truth Twitter data}

Twitter has recently released a large collection of tweets of the state--sponsored troll accounts as part of Twitter's election integrity efforts\footnote{\url{https://about.twitter.com/en_us/values/elections-integrity.html}}. We requested the unhashed version which consists of the tweets of Twitter accounts identified as Russian, Iranian and Venezuelan -- 25 million tweets shared by 8,275 troll accounts (see Table \ref{tab:groundtruth}). These troll IDs served as ground--truth identifiers of the troll users in our tweets collection.

\begin{table}[hbt]
	\centering
	\caption{Ground--truth data}\label{tab:groundtruth} 
	\setlength\tabcolsep{10pt}
	\begin{tabular}{l | l || l| l }
		\multicolumn{2}{c||}{Source: 8,275 troll accounts }     &  \multicolumn{2}{c}{Target} \\          
		\multicolumn{2}{c||}{Total tweets: 25,076,853} & trolls & real users \\
		\hline
		Replies  & 1,549,742         & 2352      &  410,779 \\
		Retweets & 8,617,208         & 3159      &  531,374 \\
		Mentions & 10,641,427        & 2885      & 1,661,716\\	 
		\hline		
	\end{tabular}	                                  
\end{table}

We observe that the majority of trolls' actions were retweets and mentions. Moreover, the target users were mostly real users, namely, Twitter accounts that were not in the ground truth troll IDs list. As we describe in the next section, we see the tweet--replies and tweet--mentions as \textit{actions}, i.e. they represent the interactions from the users who performed these actions to the users who received them. For instance, when a user $i$ replies to a tweet of user $j$ then the \textit{source} is $i$ while the \textit{target} is $j$. Hence, in Table \ref{tab:groundtruth} with the term \textit{source} we are referring to the 8,275 troll users. 

\subsection{Our Twitter dataset}
The analysis in this paper is based on 152.5 million tweets from 9.9 million users. The tweets were downloaded using the Twitter streaming (1\%) API in the period before and up to the 2016 US presidential election -- from September 21 to November 7, 2016 (47 days). The tweets' track terms were related to political content such as ``hillary2016'', ``clinton2016'', ``paul2016'', ``trump2016'' and ``donaldtrump2016''(see Appendix, Table \ref{tab:track_terms} for the complete list of the track terms). Namely, a list of phrases used to determine which Tweets are delivered by the stream (see \footnote{\url{https://developer.twitter.com/en/docs/tweets/filter-realtime/guides/basic-stream-parameters.html}} for more details). These were collected using a Python script utilizing the Tweepy module. In addition to the tweet text, user screen name, and user ID, we also collected metadata including the hashtags and expanded URL data from Twitter, information on the account creation, user timezone, and user-supplied location and biographic information.

\begin{table}[hbt]
	\centering
	\caption{The tweets collection}\label{tab:ourdata}
	\setlength\tabcolsep{18pt}
	\begin{tabular}{l | l | l }
		&    real-users  & trolls   \\         
		\hline
		
		user-IDs     &  9,939,698     & 822    \\
		Total tweets &  152,479,440   & 35,489  \\
		\hline\hline
		Replies      &  12,942,628    & 160     \\
		Mentions     &  172,145,775   & 33,627  \\	 
		Retweets     &   N/A          & N/A      \\	
		\hline
		
	\end{tabular}	                                  
\end{table}

Using the troll ground--truth IDs we identified 35.5K tweets from 822 troll accounts (see Table \ref{tab:ourdata}). 
 
\section{Methodology} \label{sec:methods} 

\subsection{The graph of interactions}
In this paper, we followed a graph theoretical approach, namely we map users to nodes and we map the interactions between users to edges. We construct the graph based on the tweets collection we presented in the previous section -- 152.5 million tweets collected during 47 days. The actions between the users are either replies or mentions (we had not collected the retweets). Each directed edge $(i, j)$ corresponds to a tweet--action from user $i$ to user $j$; either user $i$ had replied to a tweet of user $j$ or he had mentioned user $j$ in his tweet, or both. So, we distinguish between replies and mentions. Both are actions from one user to another. Hence, we have the graph of users' interactions which is a directed multigraph (i.e., multiple edges are permitted between any pair of nodes) consisting of:
\begin{itemize}
	\item nodes: (i) 821 trolls; (ii) 9,321,061 real users. 
	\item edges: 169,921,921.
	\item 31,660 edges: from 659 trolls to 9,371 real users.
	\item 670,605 edges: from 121,924 real users to 285 trolls.
	\item ego--net nodes: 127,192 real users (based on both the in--edges and out--edges with the trolls)
\end{itemize}

\begin{figure*}[htb]
	\begin{subfigure}{.5\textwidth}
		\centering
		\includegraphics[width=.8\linewidth]{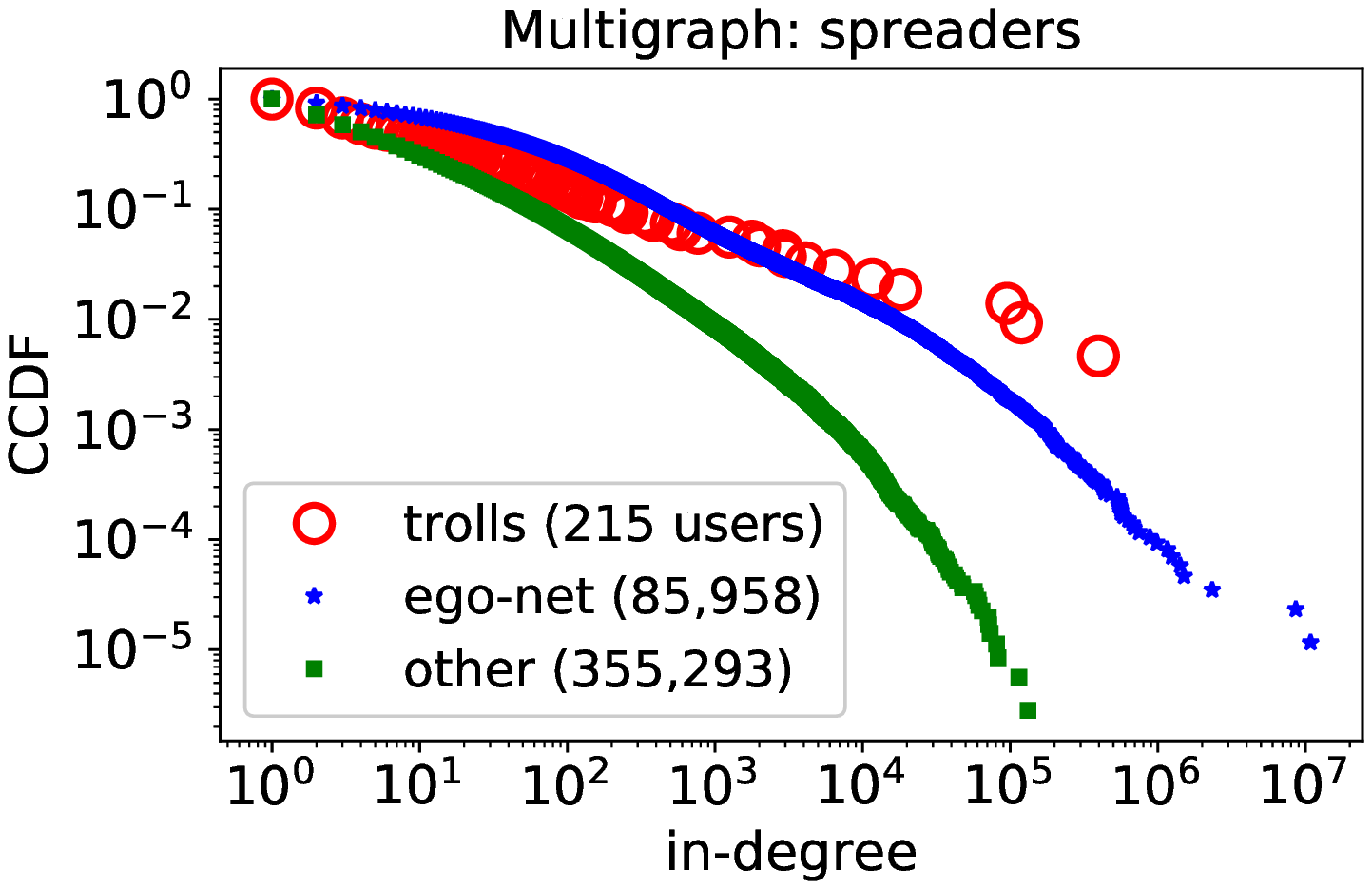}  
		\caption{}
		\label{}
	\end{subfigure}
	\begin{subfigure}{.5\textwidth}
		\centering
		\includegraphics[width=.8\linewidth]{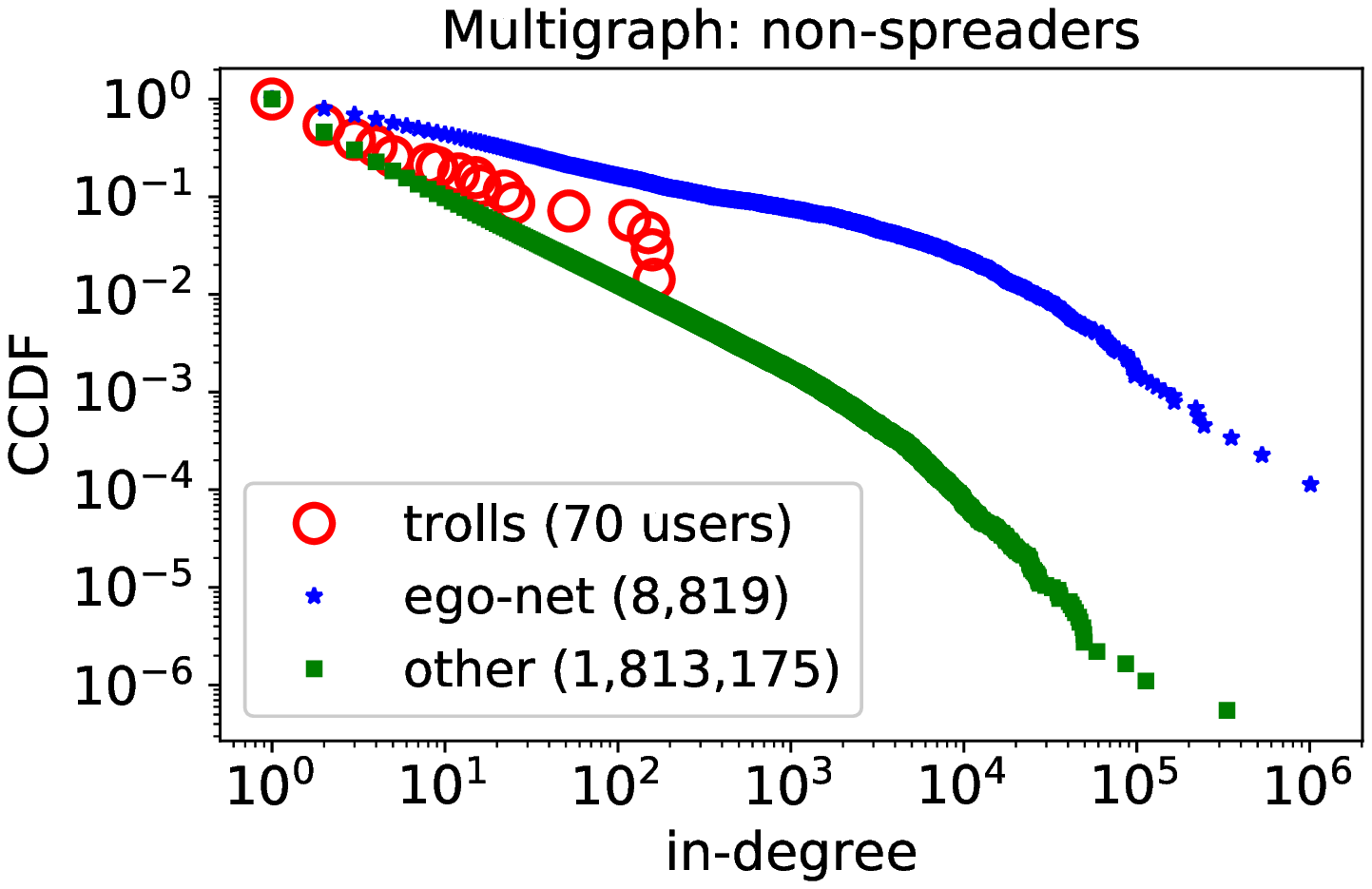}  
		\caption{}
		\label{}
	\end{subfigure}
	\begin{subfigure}{.5\textwidth}
		\centering
		\includegraphics[width=.8\linewidth]{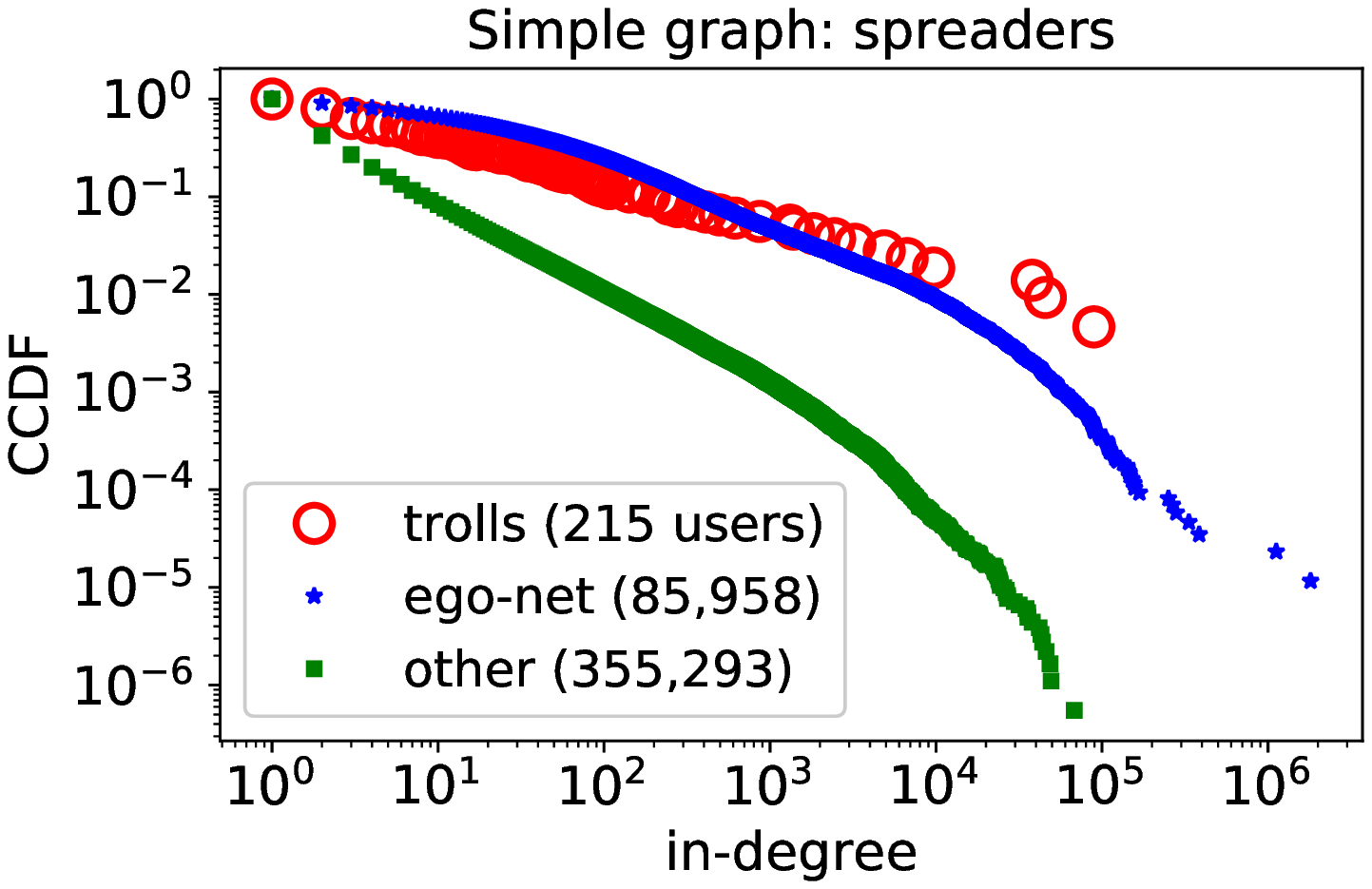}  
		\caption{}
		\label{}
	\end{subfigure}
	\begin{subfigure}{.5\textwidth}
		\centering
		\includegraphics[width=.8\linewidth]{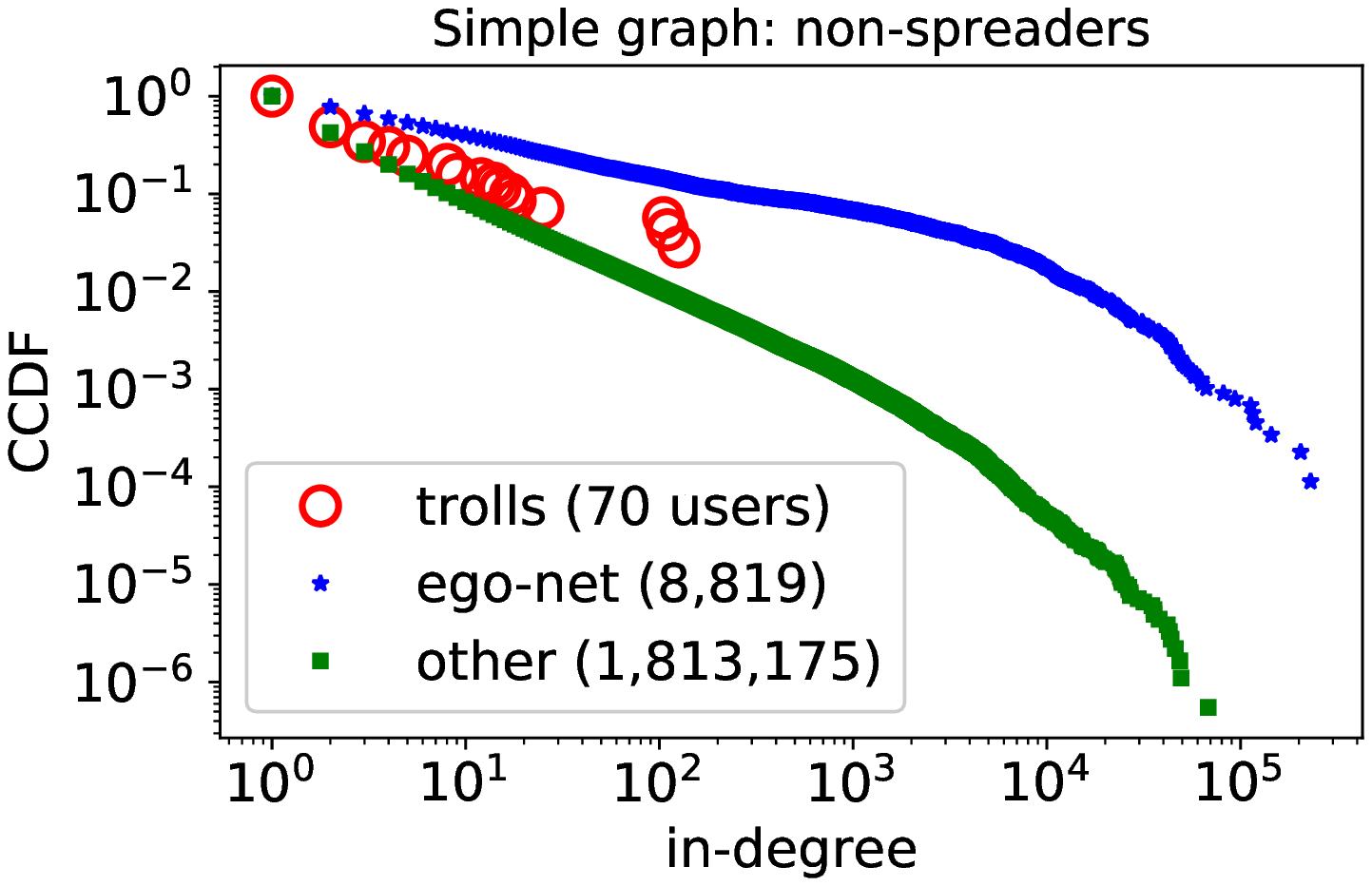}  
		\caption{}
		\label{}
	\end{subfigure}
	\caption{Multigraph \& simple graph: CCDF of the in--degree for spreaders and non--spreaders.}
	\label{fig:in-degree}
\end{figure*}

\begin{figure*}[ht]
	\begin{subfigure}{.5\textwidth}
		\centering
		\includegraphics[width=.8\linewidth]{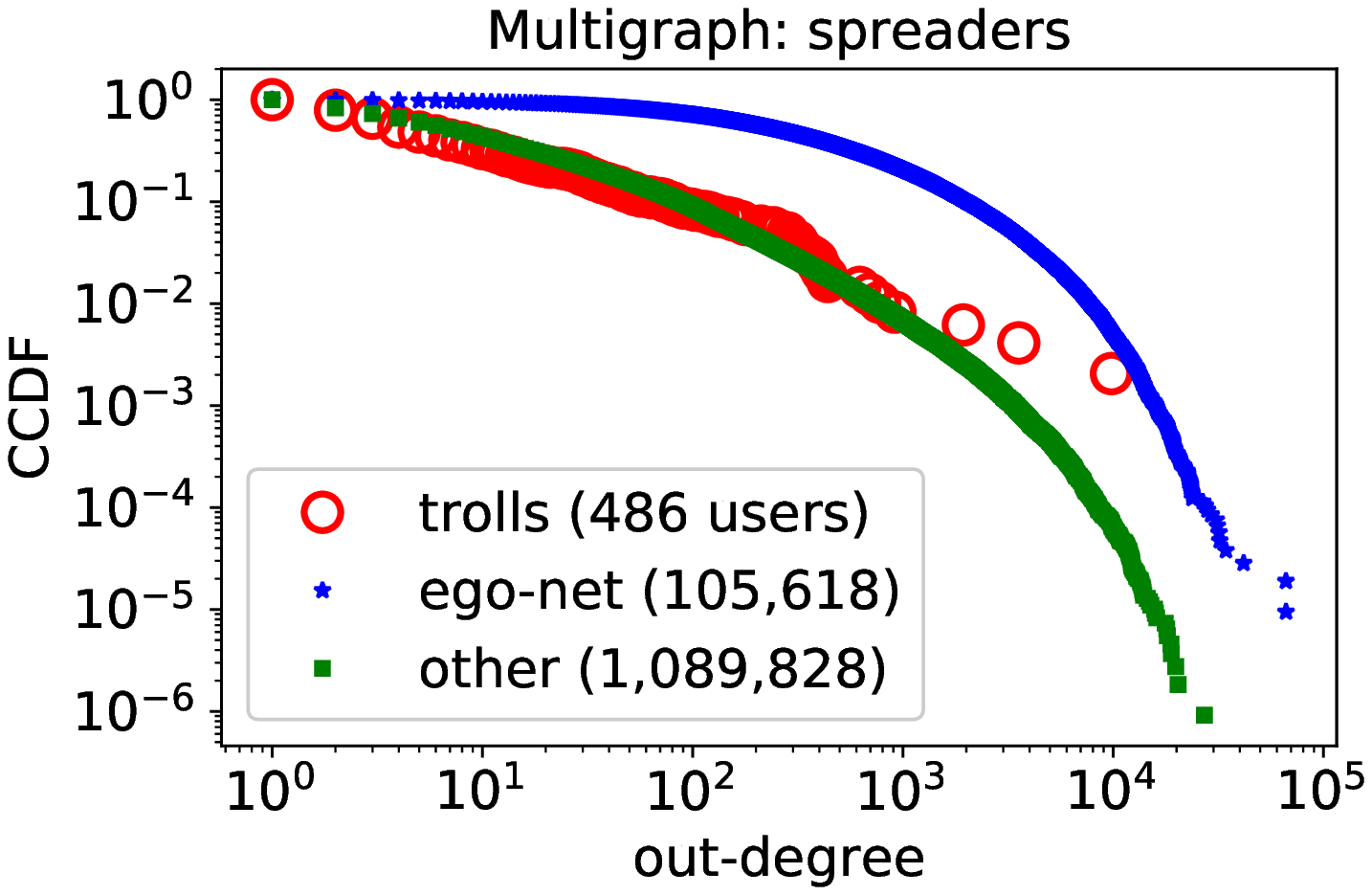}  
		\caption{}
		\label{}
	\end{subfigure}
	\begin{subfigure}{.5\textwidth}
		\centering
		\includegraphics[width=.8\linewidth]{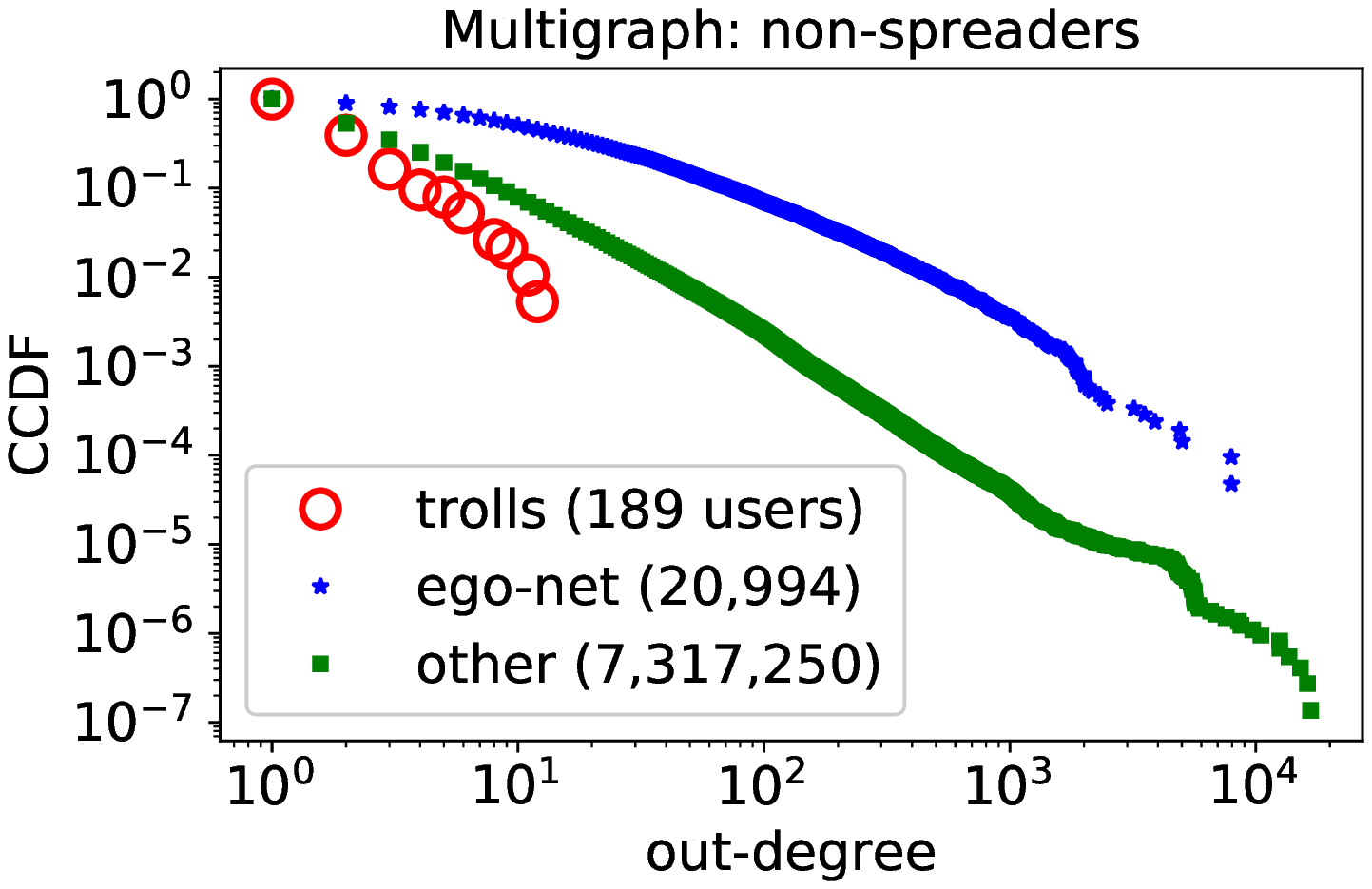}  
		\caption{}
		\label{}
	\end{subfigure}
	\begin{subfigure}{.5\textwidth}
		\centering
		\includegraphics[width=.8\linewidth]{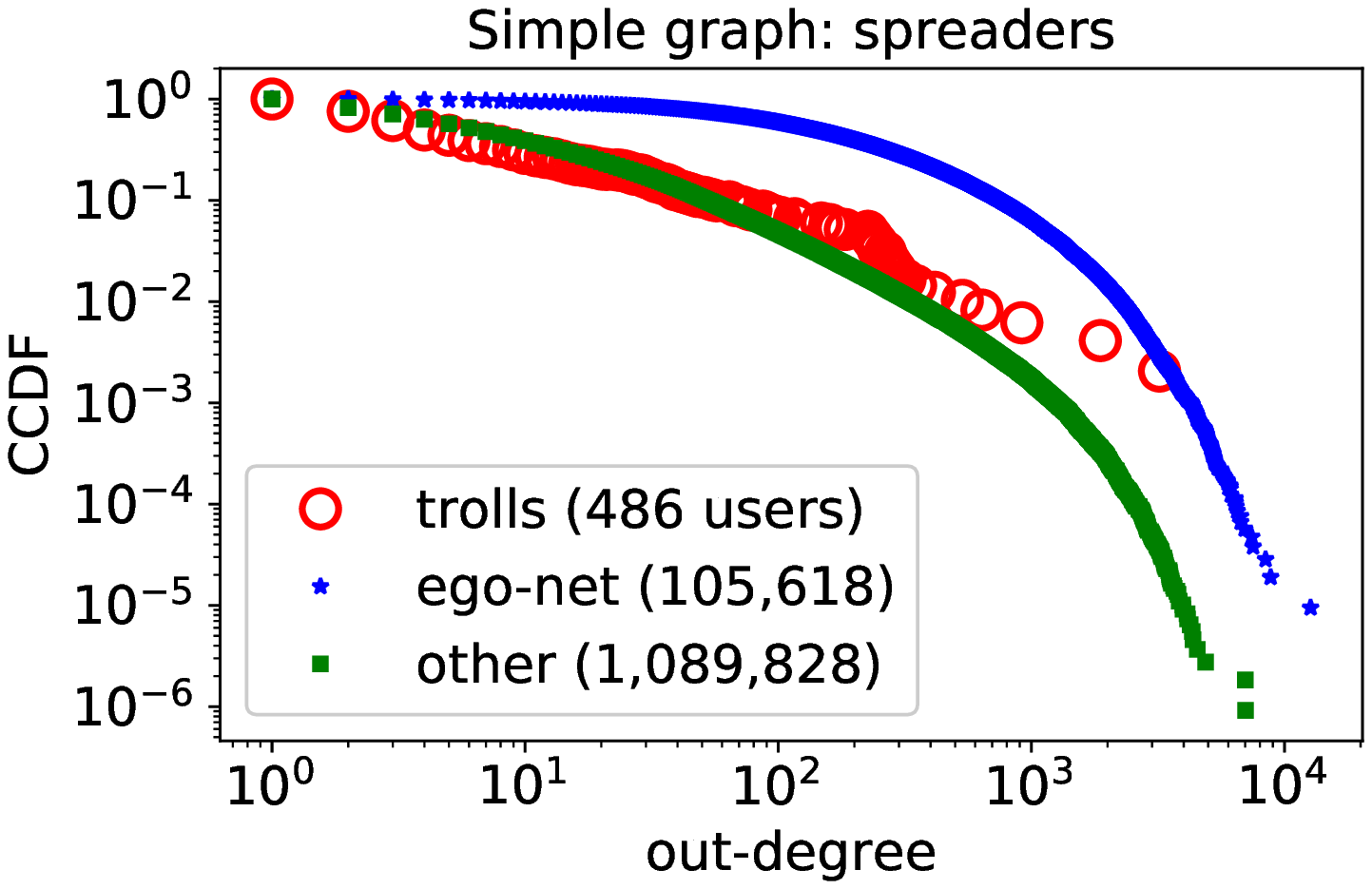}  
		\caption{}
		\label{}
	\end{subfigure}
	\begin{subfigure}{.5\textwidth}
		\centering
		\includegraphics[width=.8\linewidth]{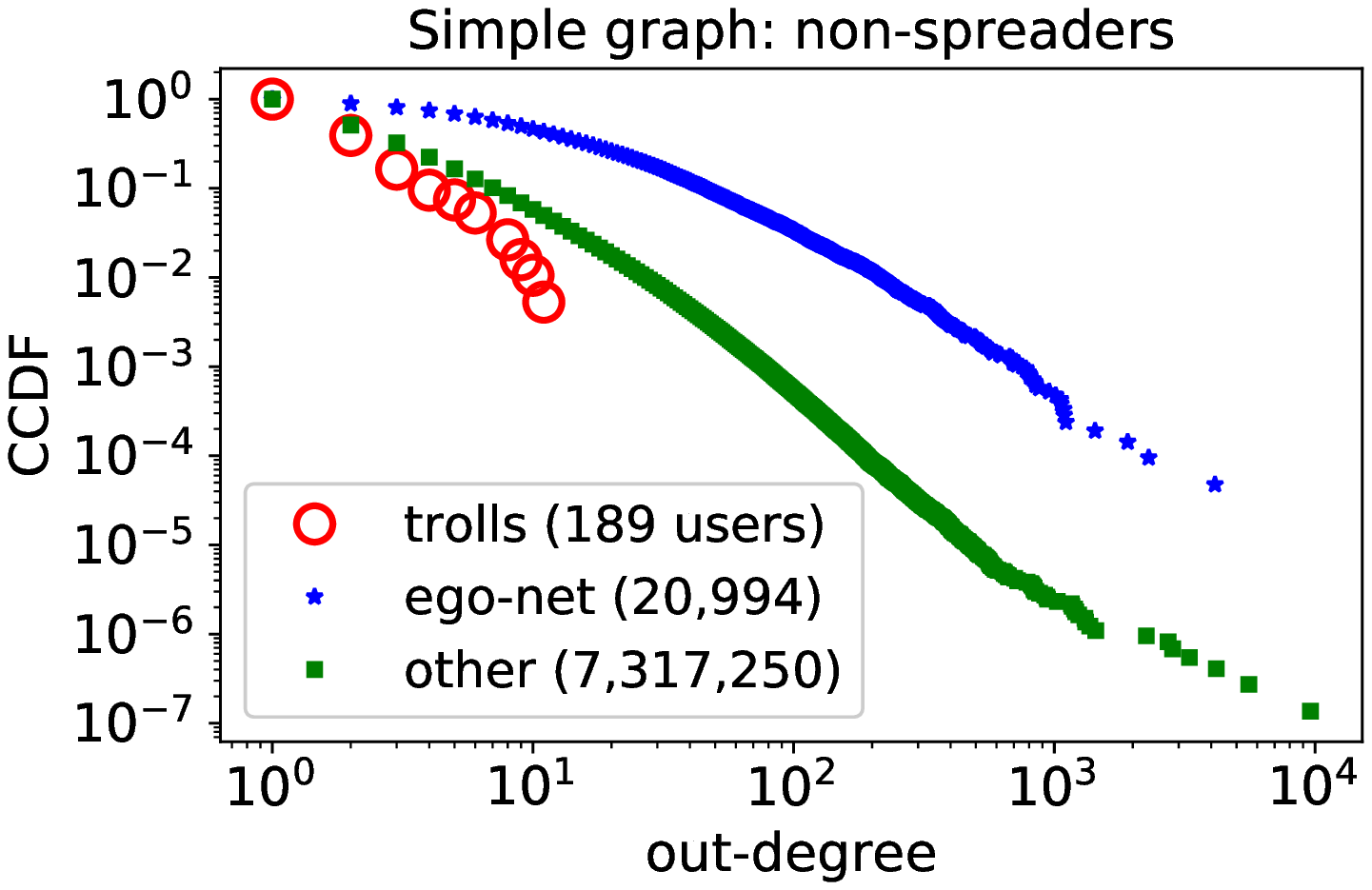}  
		\caption{}
		\label{}
	\end{subfigure}
	\caption{Multigraph \& simple graph: CCDF of the out--degree for spreaders and non--spreaders.}
	\label{fig:out-degree}
\end{figure*}

\begin{figure*}[htb]
	\begin{subfigure}{.5\textwidth}
		\centering
		\includegraphics[width=.8\linewidth]{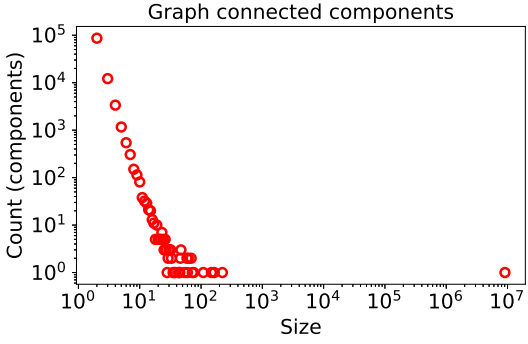}  
		\caption{}
		\label{}
	\end{subfigure}
	\begin{subfigure}{.5\textwidth}
		\centering
		\includegraphics[width=.8\linewidth]{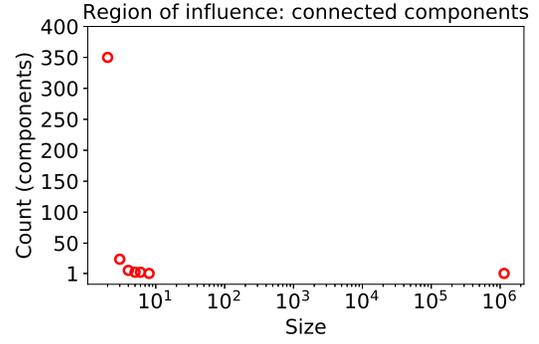}  
		\caption{}
		\label{}
	\end{subfigure}
	\caption{The connected components of the undirected versions: (a) simple graph; (b) region of influence}
	\label{fig:components}
\end{figure*}

\begin{figure*}[htb]
	\begin{subfigure}{.5\textwidth}
		\centering
		\includegraphics[width=.8\linewidth]{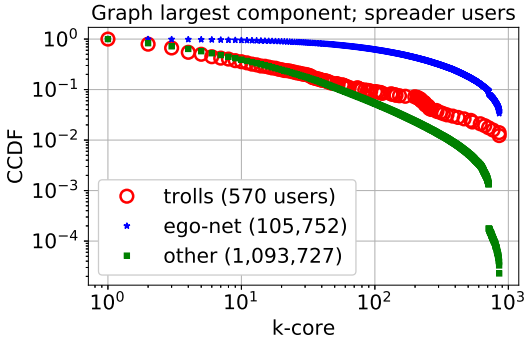}  
		\caption{}
		\label{}
	\end{subfigure}
	\begin{subfigure}{.5\textwidth}
		\centering
		\includegraphics[width=.8\linewidth]{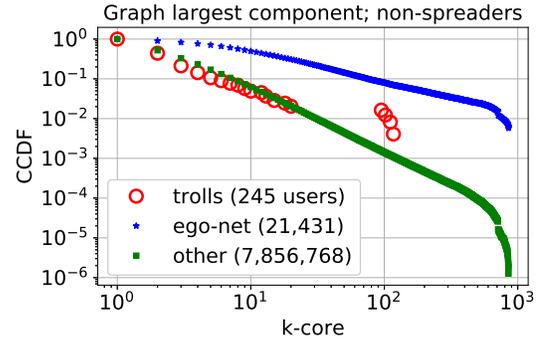}  
		\caption{}
		\label{}
	\end{subfigure}
	\caption{CCDF of the k-core values for the nodes in the largest connected component.}
	\label{fig:kcore}
\end{figure*}

\begin{figure}[htb]
	\centering
	\includegraphics[width=.85\linewidth]{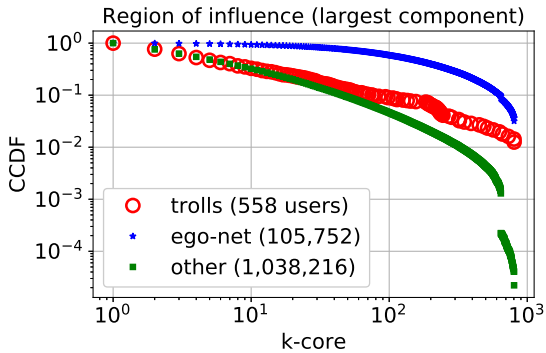} 
	\caption{Region of influence, largest connected component: CCDF of k-core values}
	\label{fig:kcore_region}
\end{figure}

\begin{figure*}[htb]	
	\begin{subfigure}{.5\textwidth}
		\centering
		\includegraphics[width=.8\linewidth]{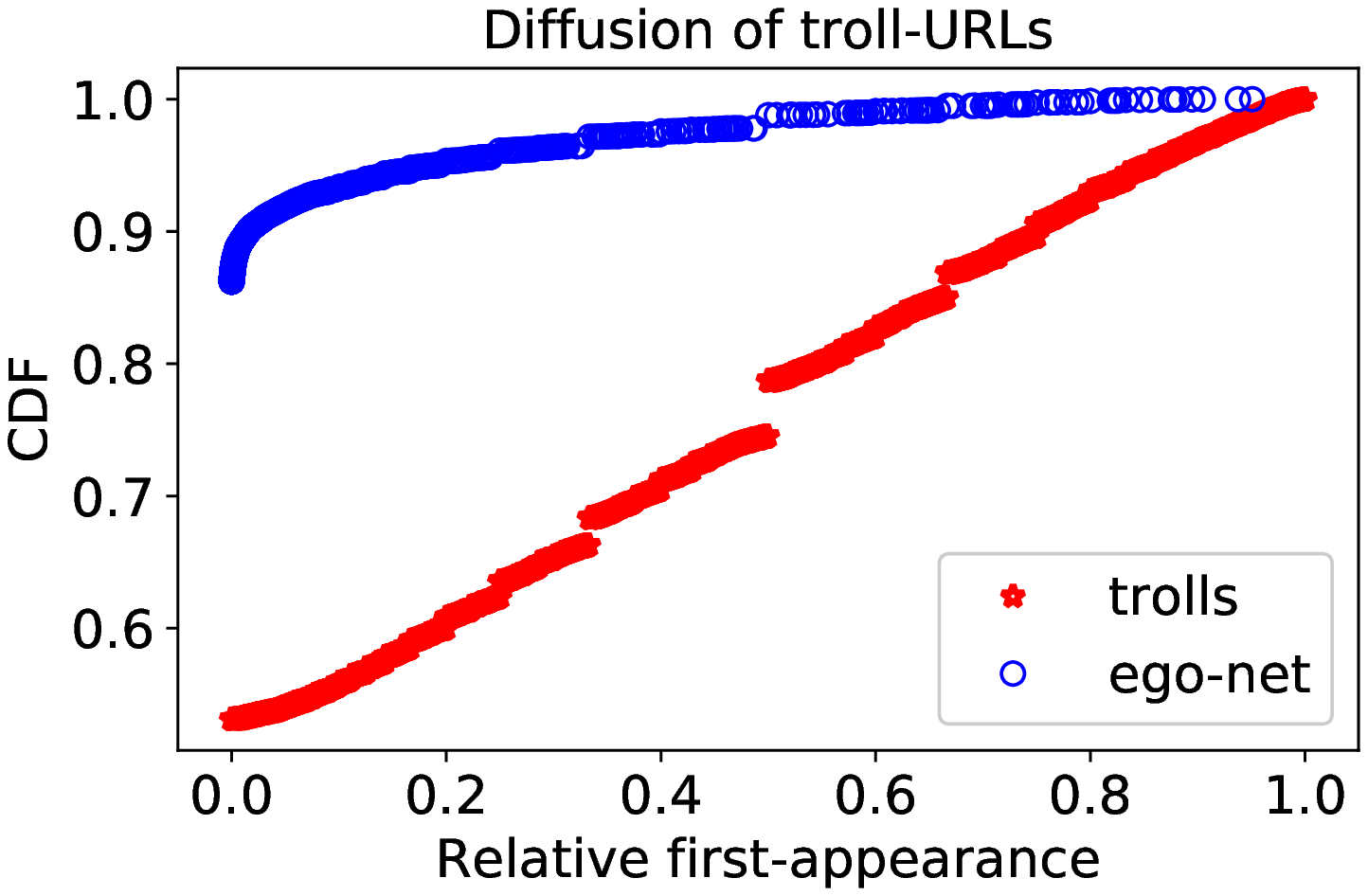}  
		\caption{}
		\label{}
	\end{subfigure}
	\begin{subfigure}{.5\textwidth}
		\centering
		\includegraphics[width=.8\linewidth]{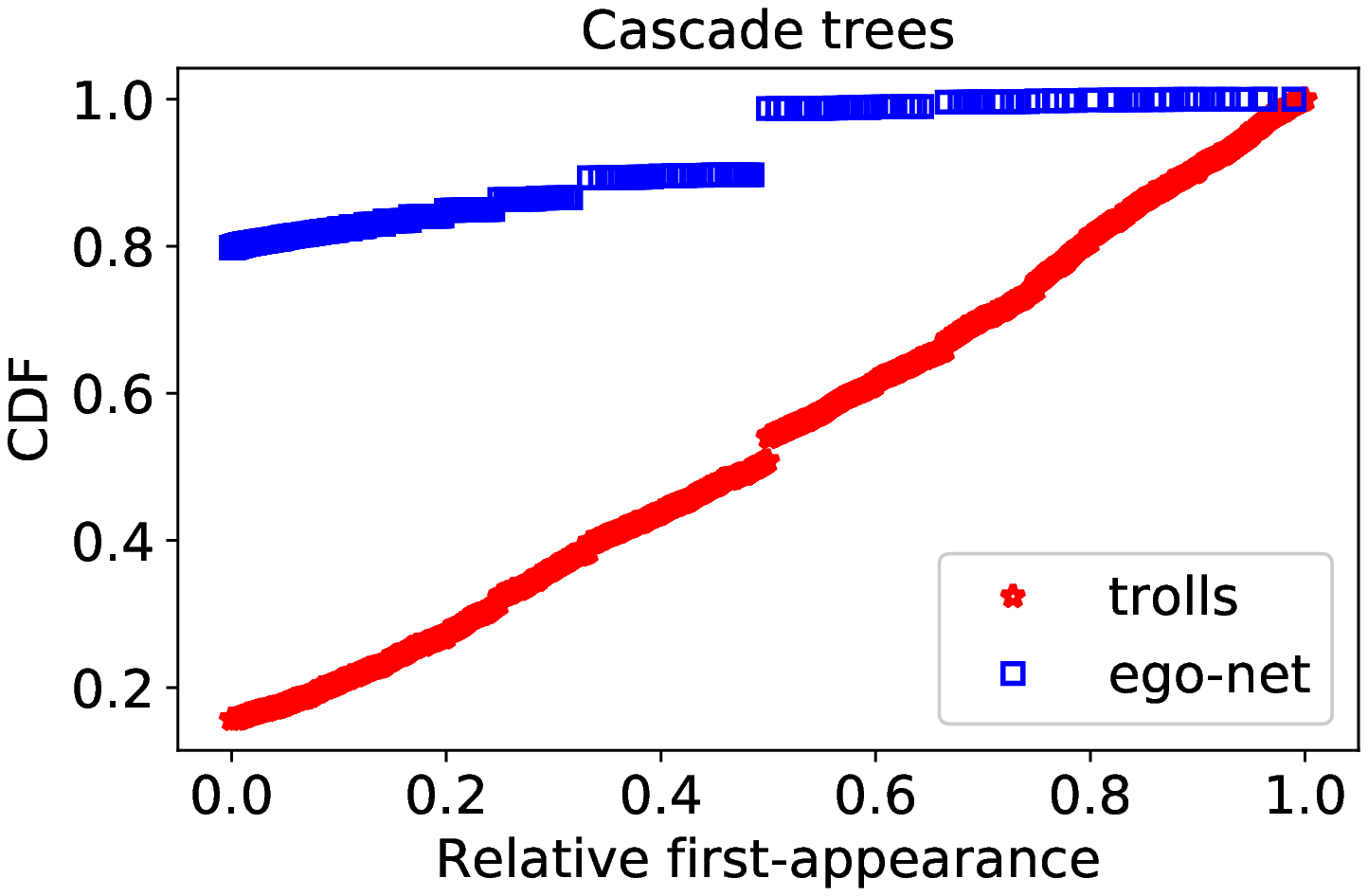}  
		\caption{}
		\label{}
	\end{subfigure}
	\begin{subfigure}{.5\textwidth}
		\centering
		\includegraphics[width=.8\linewidth]{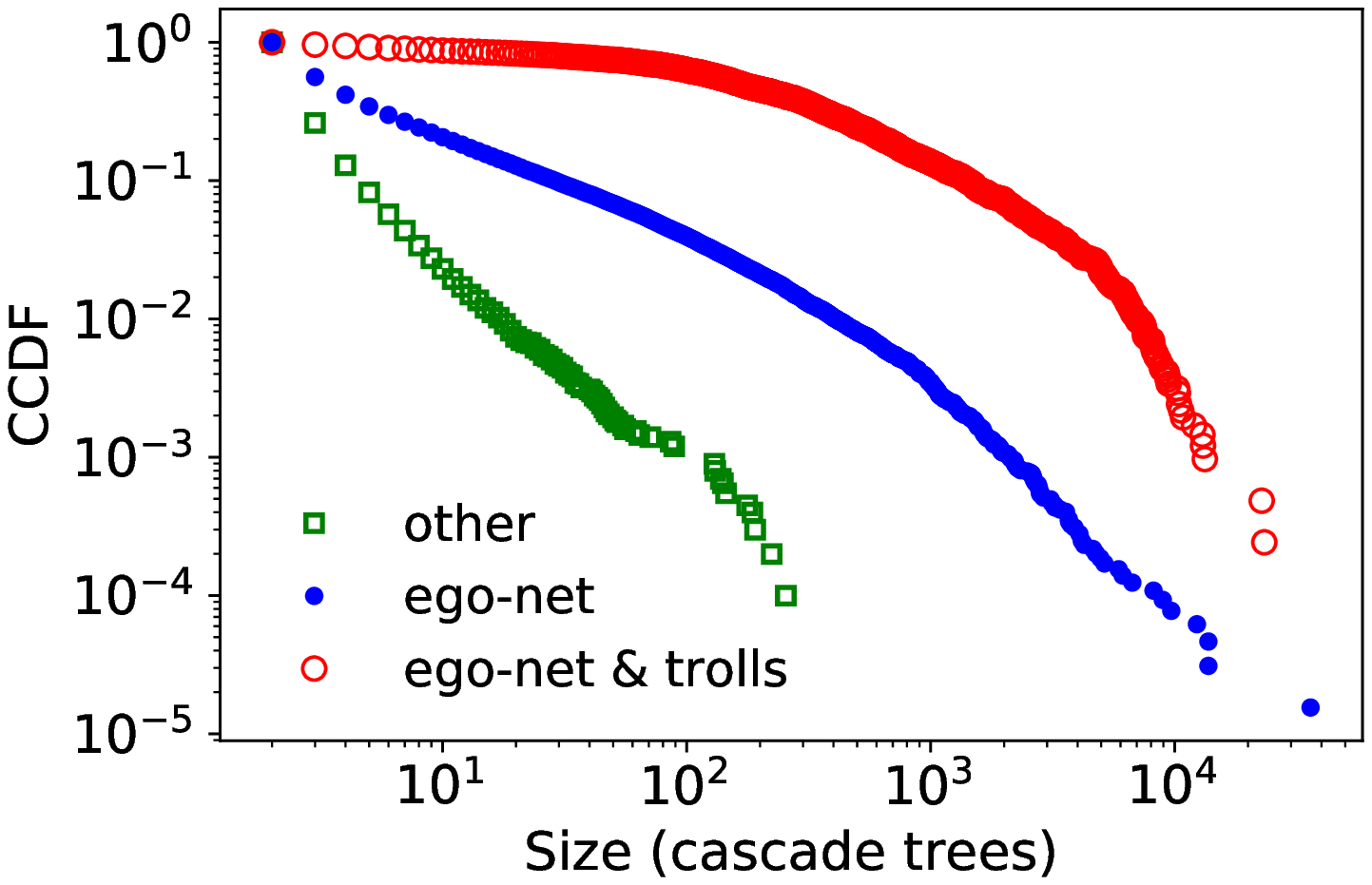}  
		\caption{}
		\label{}
	\end{subfigure}			
	\begin{subfigure}{.5\textwidth}
		\centering
		\includegraphics[width=.8\linewidth]{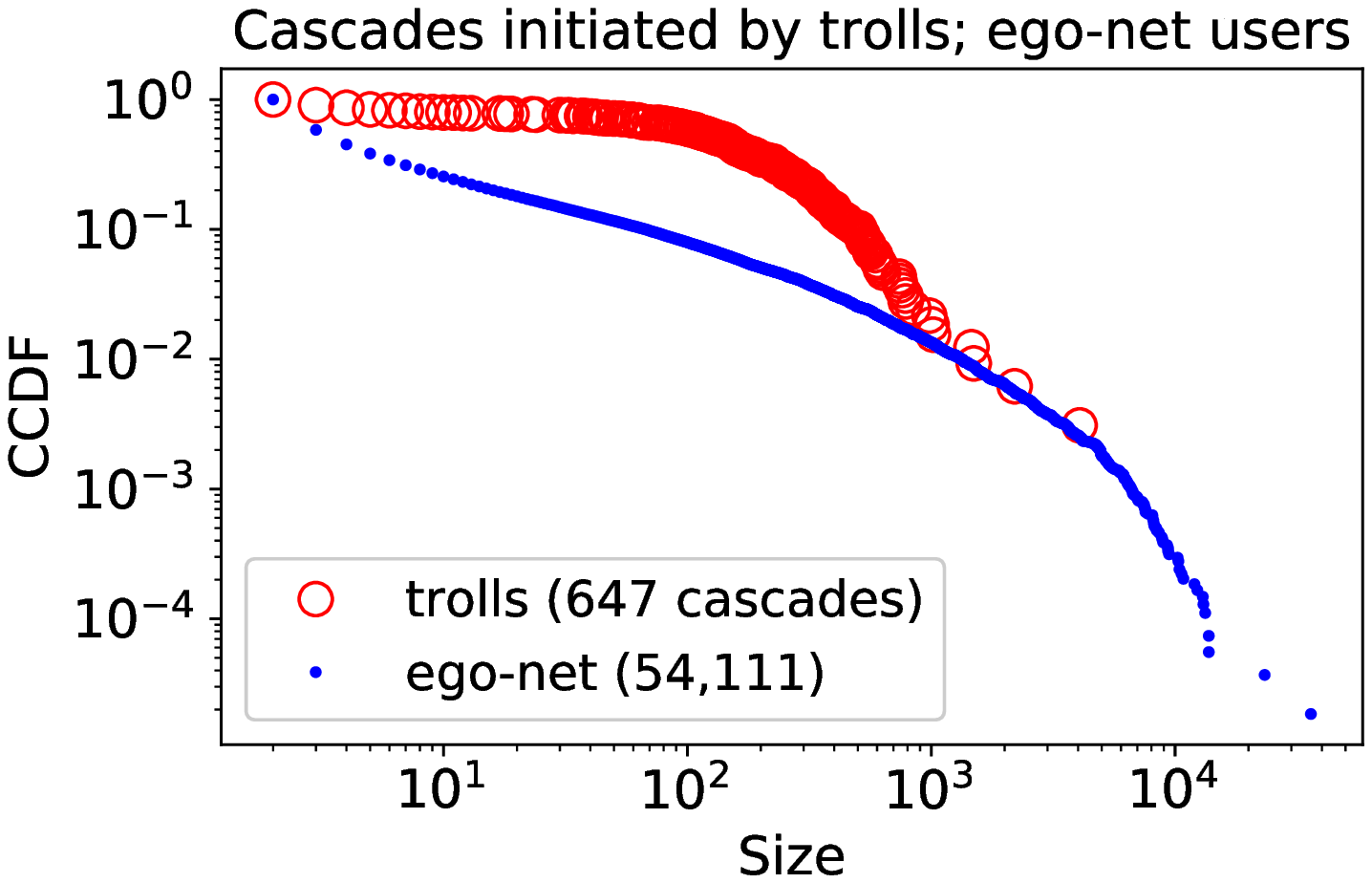}  
		\caption{}
		\label{}
	\end{subfigure}							
	\caption{URL--diffusions with more than 100 distinct users: (a) CDF of the relative first--position in the URL--diffusions; (b) The relative first--position of the trolls as well as the ego--net users in the cascade trees; (c) CDF of the cascades size for the cascades where: i) at least one troll and one ego--net user are present, ii) at least one ego--net user and no trolls, iii) other cascades; (d) CCDF of the cascades size initiated by trolls and ego--net users.}
	\label{fig:cascades}
\end{figure*}

\begin{figure*}[htb]		
	\begin{subfigure}{.5\textwidth}
		\centering
		\includegraphics[width=.8\linewidth]{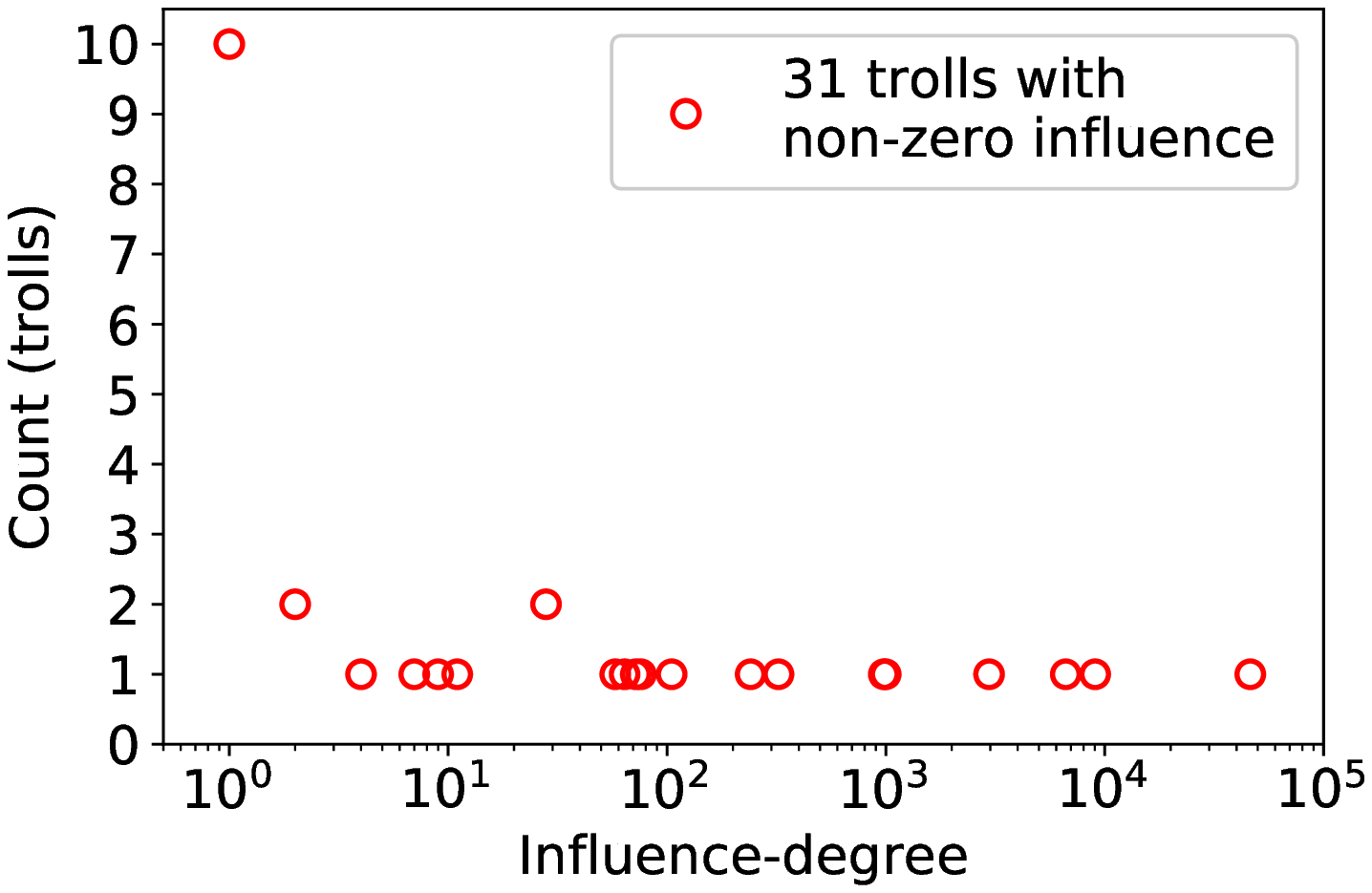} 
		\caption{}
		\label{}
	\end{subfigure}
	\begin{subfigure}{.5\textwidth}
		\centering
		\includegraphics[width=.8\linewidth]{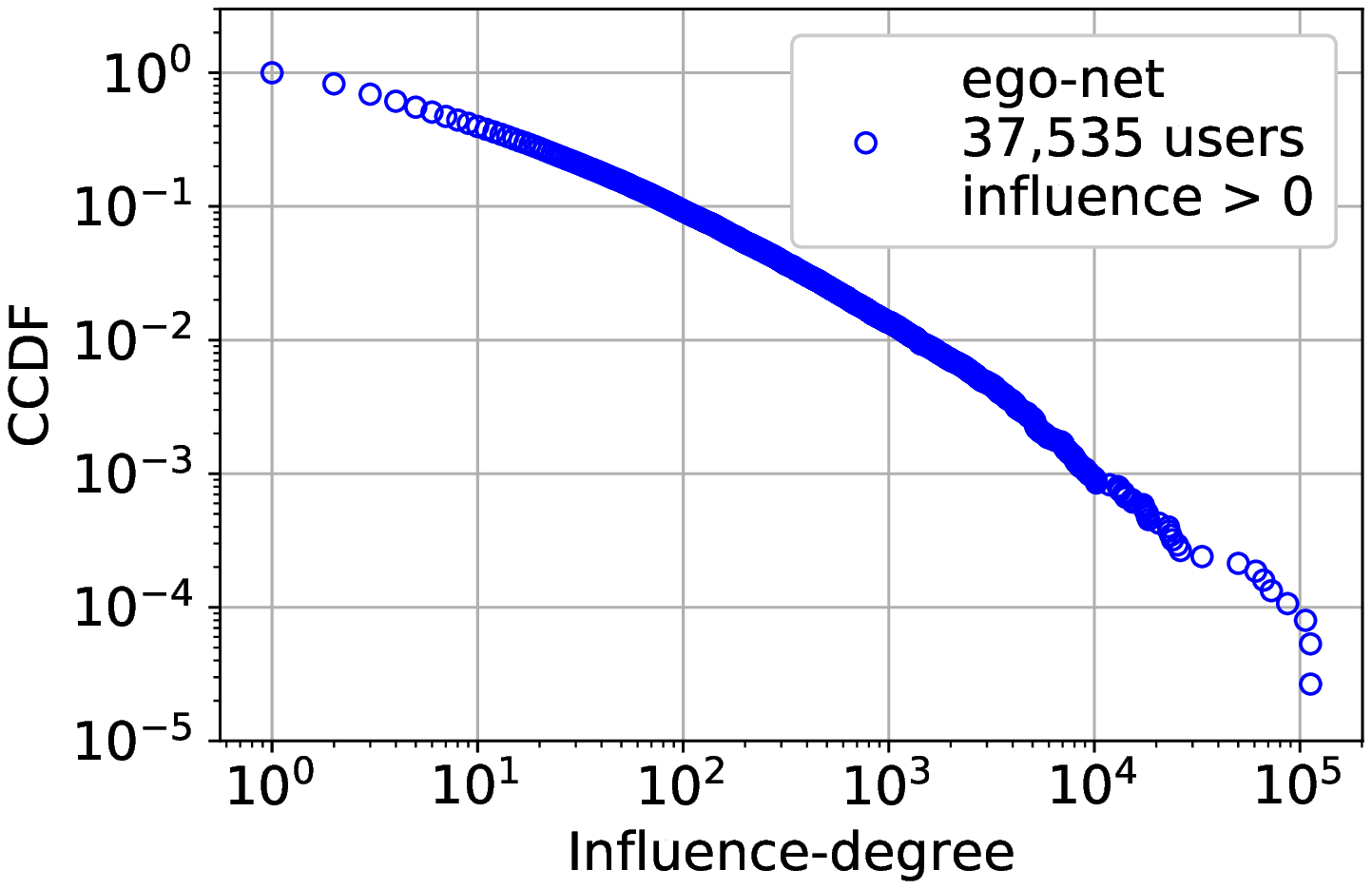} 
		\caption{}
		\label{}
	\end{subfigure}
	\begin{subfigure}{.5\textwidth}
		\centering
		\includegraphics[width=.8\linewidth]{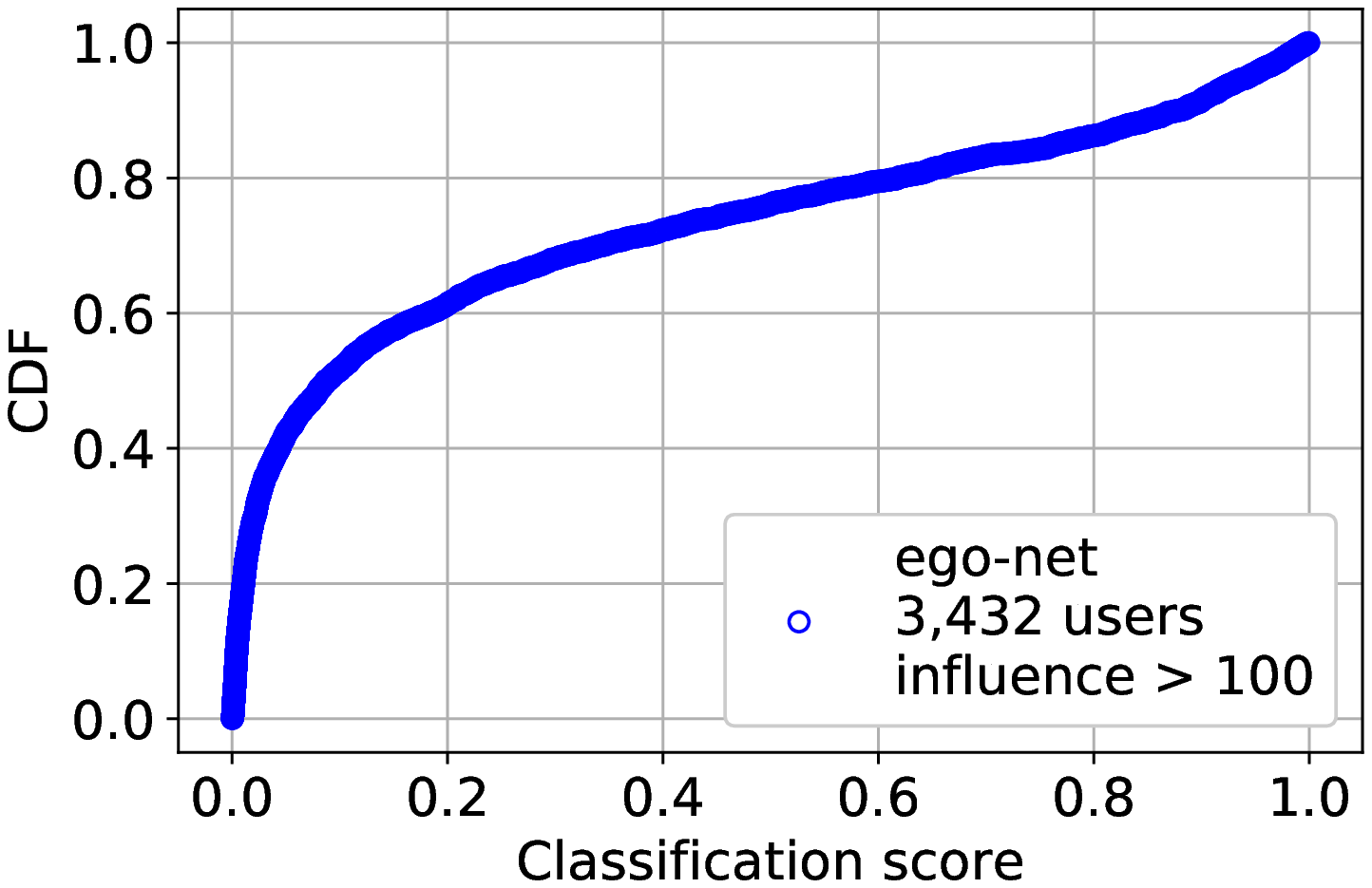}  
		\caption{}
		\label{}
	\end{subfigure}
	\begin{subfigure}{.5\textwidth}
		\centering
		\includegraphics[width=.8\linewidth]{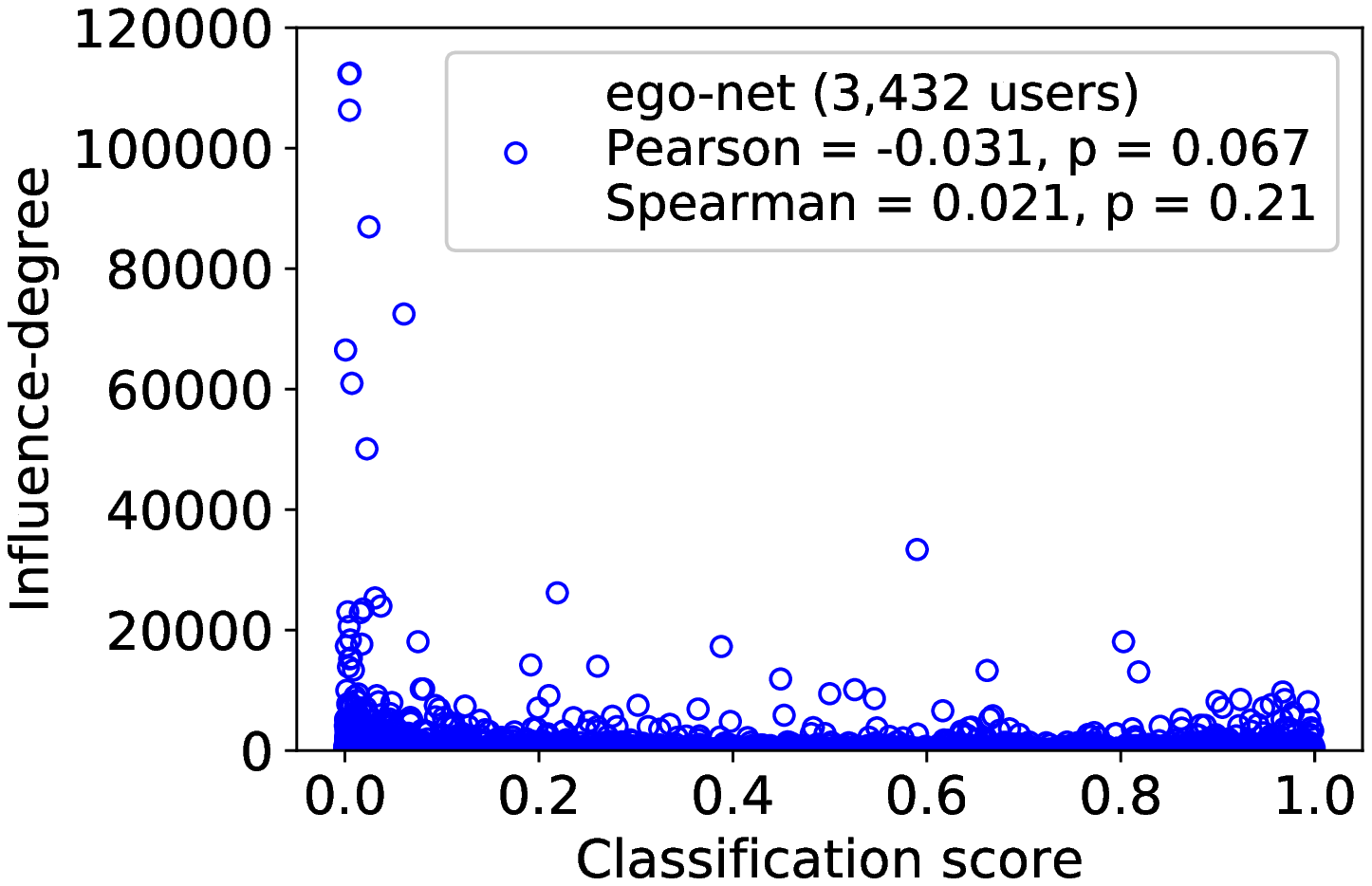}  
		\caption{}
		\label{}
	\end{subfigure}    
	\caption{Influence: (a) Number of trolls with a given influence--degree value (for the troll users with with $\text{influence--degree} > 0)$ ; (b) CCDF of the $\text{influence--degree}$ for the ego--net users with non--zero influence values; (c) CCDF of the classification scores for the ego--net users with $\text{influence--degree} > 100$; (d) Scatter plot of the classification scores and the influence--degree for the ego--net users with $\text{influence--degree} > 100$.}
	\label{fig:influence}
\end{figure*}

\begin{figure}[htb]
	\centering
	\includegraphics[width=.85\linewidth]{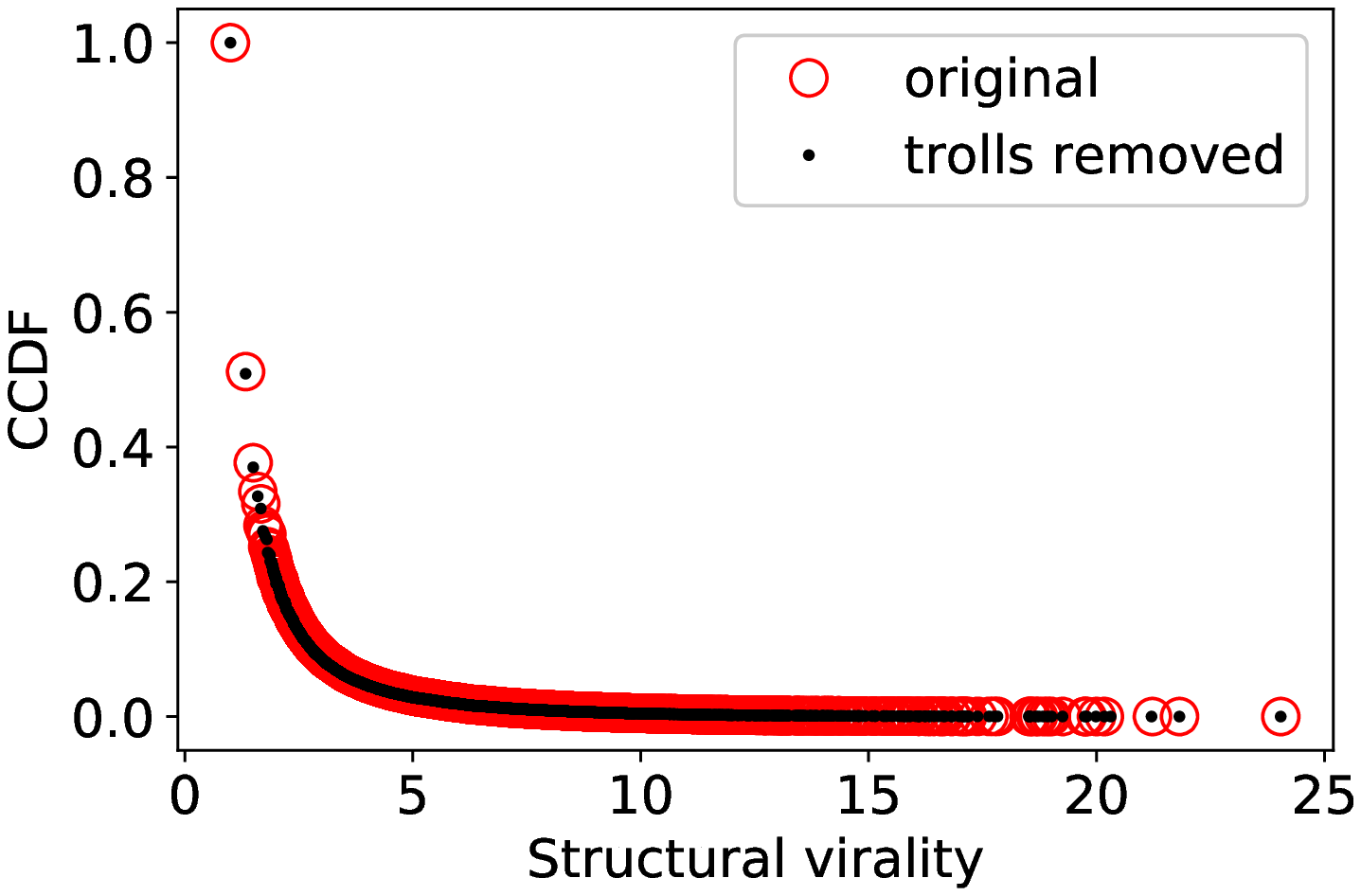} 
	\caption{The structural virality of the original cascade trees as well as the cascade trees when the trolls have been removed from the multigraph graph.}
	\label{fig:virality}
\end{figure}

\subsection{The region of influence}

The main goal of this study is to quantify the influence trolls had over real (authentic) users. A challenge we have to tackle is that only few trolls appear in our dataset and the actions/edges from trolls to real users are small in size. Moreover, we have 121,924 real-users which have at least one action with trolls (670,605 total actions). These users have either replied to troll tweets or they have mentioned a troll ID in their tweets. These real users are the troll ego-nets, namely the trolls' nearest neighbors -- one--hop distance from the trolls in the multigraph. Yet, this number is not considerably large compared to the millions of total actions. Therefore, we have to expand the possible region of users by identifying a population that might have been influenced by troll activities. In order to achieve that, we use the web--URLs and media--URLs that have been shared by the trolls as an identifier of troll influence anywhere in the graph. 

Specifically, we collect the \textit{expanded URL}, \textit{media expanded URL} and \textit{media URL https} from all tweets. Then, we concentrate our attention to the URLs that have been shared by at least one troll user. For the rest of the paper we call this set \textit{troll--URLs}. By that, we do not imply that these web contexts necessarily belong to trolls. They are just identifiers of the trolls' strategies, materials that the trolls wish to be viral. In other words, even if they are authentic web materials, they still reflect the political and strategic objectives of the trolls' disinformation campaigns.

We have 5,876,674 tweets from 620 trolls and 1,213,037 real users. These tweets contain: (i) 16,261 expanded URLs; (ii) 7,266 media expanded URLs; (iii) 7,259 media URL https.

We have three groups of users that shared troll-URLs: (i) the trolls; (ii) the real users that are part of the trolls' ego--nets; and (iii) the real users that are not directly connected with any of the trolls users. For the rest of the paper, we call all these users \textit{spreaders}.

Then, we construct the directed induced subgraph based on this set of users, namely we select only the edges between the users that appear in the directed simple graph version of the multigraph. This subgraph represents the \textit{region of influence} between trolls and real users, namely the largest region in the graph where the influence between trolls and real users could have taken place. In summary, we have: (i) nodes: 1,145,363; (ii) edges: 46,429,227; (iii) 560 trolls; (iv) 105,755 ego--net users; (v) 1,039,552 other users.

Hence, our goal is to compare the impact of these groups in the diffusion cascades of the trolls--URLs; the media and web context, which, as we mentioned previously, are a representation of the trolls disinformation objectives. The rationale of this approach is that if the trolls were the important factor in these diffusion cascades then their role during the 2016 US presidential election was substantial. Otherwise, the real users where responsible for the virality (if there was any) of the trolls' materials.

\subsection{Diffusion Cascades}
In order to compare the influence of trolls with that of real users, we have to reconstruct the path of information flow. Hence, for each troll--URL we construct the \textit{time--inferred diffusion cascades} in a manner based on a causality assumption that has been previously used in the literature \cite{Goel2015:virality,Vosoughi:2017,Vosoughi:2018}. 

\subsubsection{Causality hypothesis}
We construct the \textit{time--inferred diffusion cascades} for every troll--URL as follows: \\
Suppose that the URL $x$ has been shared by user $i$ on date $t_{i}$. First, we select the users with whom $i$ was connected with in the graph before the date $t_{i}$, i.e., there were graph edges from $i$ to them. In other words, we select the ego--net of $i$ for $t < t_{i}$. If this set is not empty, then we examine whether any of these users had also shared the URL $x$ on a date $t < t_{i}$. Suppose that two users $j$ and $k$ had shared URL $x$ on dates $t_{k} < t_{j}$. Then, we assume that user $i$ had been influenced by user $j$ on adopting and diffusing the URL $x$ and we update the cascade tree with the edge $(j, i)$. 

This approach has been previously used in the literature mainly for constructing retweet cascade trees based on followers--followees original graph. Since a user can retweet a tweet only once, there is no repetition in the user actions and the size of a cascade corresponds to the number of unique users involved in that cascade. In our case this is not true. The user actions may correspond to several tweets, hence repetitions are possible. In order to deal with this case, we assume that for a given URL, every user can be influenced only once by another user, the first time that he shared it.  

\subsection{Classification of users} 
We collect all the URLs and media URLs (the ``troll--URLs'') that were included in trolls tweets. 
Then, in our tweets collection, we identify all the tweets that have at least one troll--URL and have been shared either 
by trolls or by real users. So, our goal here is to compute a similarity measure between the troll accounts and the real 
users based on the context of their tweets. In this way, we extend the graph analysis in order to further characterize the 
interactions between the two groups of users, as well as the source of the viral cascades. For instance, if the very active and influential real users have high similarity with the trolls, then it is possible that these users were also trolls. To this end, we 
take the following steps:

\begin{enumerate}
	\item We extract the textual data and train a \textit{Doc2Vec}~\cite{Quoc2014} model.
	\item For each tweet, we get a fix-length vector to represent its content.
	\item For each user, we generate a sequence with time information and textual information.
	\item We feed sequences of users to a \textit{Bi-LSTM}~\cite{GravesS2005} model to classify them into different categories (3 categories in our setting: 1. Trolls; 2. The real-users that are connected with the trolls in the action graph; and
	3. The rest of the real-users).
	\item Due to the imbalance of size among the three groups, we randomly select 620
	users from each group to form our dataset. $2/3$ of them are used for training and the rest are used for testing.
\end{enumerate}

\begin{table}[hbt]
	\centering
	\caption{The classifier}\label{tab:classifier} 
	\setlength\tabcolsep{3pt}
	\begin{tabular}{l | l | l| l |l }
		class & precision  &  recall & f1-score &  support \\ \hline
		
		trolls (620)           &   0.8009   &  0.8204   &  0.8106   &   206 \\
		ego--net (105,755)           &   0.9259   &  0.6039   &   0.7310  &   207 \\
		others (1,107,282)            &   0.6533   &  0.8647   &  0.7443   &   207 \\
		\hline\hline
		micro avg     &   0.7629   &  0.7629   &  0.7629   &   620 \\
		macro avg     &   0.7934   &   0.7630  &  0.7619   &   620 \\
		weighted avg  &    0.7934  &   0.7629  &   0.7619  &   620 \\ 	 
		\hline		
	\end{tabular}	                                  
\end{table}

Table~\ref{tab:classifier} shows the performance of our classifier. We can see that the F1-score of trolls 
detection reaches 0.8106 with both precision and recall higher than 0.8. This illustrates the usefulness of our classifier.  

\section{Results} \label{sec:results}
The analysis is based on the comparison of the influence of three groups of users:
(i) the trolls; (ii) the real-users that are directly connected with the trolls (the trolls' ego--net); (iii) the rest of the 
real users. As aforementioned, in order to identify a broad region of users that might have been influenced by the trolls, 
we use the web and media URLs as identifiers of trolls influence. We call \textit{spreaders} the users that had shared at least one troll--URL. Hence, the three groups of users can be further divided in \textit{spreaders} and \textit{non--spreaders}.

The analysis we present in this section consists of the following steps:

\begin{enumerate}
	\item We compute the degree distribution in the overall directed multigraph as well as in the corresponding directed simple graph (where only one edge is allowed between each pair of nodes).
	\item We analyze the undirected version of the simple graph and we identify its connected components. Moreover, we compute the k-core values for the nodes in the largest connected component. We repeat this analysis for the region of influence.
	\item We compute the classification scores of the users in the region of influence. Based on these scores, we estimate the similarity between trolls and real users.
	\item We analyse the characteristics of the time inferred diffusion cascades of the trolls--URLs and we compute their structural vitality.            
\end{enumerate}

\subsection{Graph topology}

\subsubsection{Degree distribution}
In Figures \ref{fig:in-degree} and \ref{fig:out-degree} we present the empirical complementary cumulative distribution (CCDF) of the
in--degree and out--degree for each node/user in the directed multigraph as well as in the directed simple graph. We construct the
simple graph by allowing only one edge for each pair of nodes that is already connected in the multigraph. Moreover, the multigraph has been constructed based on the users' actions (replies and mentions) on other Twitter accounts and posts. Hence, the in--degree represents the nodes' popularity; that is, how many actions correspond to users interested in their tweets or their Twitter account in general. On the other hand, the out--degree is a measure of users' sociability/extroversion, i.e. how many actions a given user has performed on other Twitter accounts. Furthermore, it is important to compare the degree distributions in both graphs (multigraph and simple graph) because users with high degree in the multigraph do not necessarily have large degree in the simple graph. For instance, a given user might have a large in--degree in the multigraph only because he is popular to a small group of people which is highly engaged with his Twitter account -- they perform a large number of actions on the tweets of the user in question. Hence, the user will have small in--degree in the simple graph.      

As mentioned earlier, the real users can be divided in four groups based on the graph proximity they have with the trolls and the material they have shared. Hence, we have the ego--net and the rest of the users as well as the spreaders and non--spreaders. As can be seen in Figures \ref{fig:in-degree} and \ref{fig:out-degree}, the spreaders that belong to the trolls' ego--net are the most active 
and the most popular ones. Moreover, the degree distributions between multigraph and simple graph are almost identical for the four
groups of users. 

\subsubsection{Connected components}
Here we examine the structure of the undirected version of the simple graph by computing the connected components. A connected 
component is a subgraph where for each pair of nodes $i$, $j$ there is an undirected path -- a graph traverse -- from $i$ to $j$. 
Since the subject of this study is the diffusion of information, the connectivity of a region implies that there is a possible path for
information flow between the nodes that belong to this region. 

The undirected simple graph consists of 9,321,882‬ nodes and 82,842,096 edges. We identify 104,954 connected components. Figure~\ref{fig:components}(a) presents the number of connected components for a given component size (i.e. number of nodes in the component) in a log--log plot. The largest part of the graph is well connected. The largest connected component -- undirected subgraph -- consists of 9,078,493 nodes and 82,698,678 edges while the second one has only 223 nodes. Moreover, 815 trolls and 127,183 ego--net users are in the largest connected component.

Regarding the region of influence, its undirected version consists of 388 connected components. The largest connected component has 1,144,526 nodes while the second largest has only 8 nodes (see Figure~\ref{fig:components}(b)). 

\subsubsection{k--core decomposition}
We compute the \textit{$k$--core decomposition} of the nodes in the largest connected component of the undirected versions of the
overall graph and the region of influence. The k--core values is one of the most effective centrality measures for identifying the 
influential spreaders in a complex network \cite{kitsak:2010}. 

In Figures \ref{fig:kcore}(a) and \ref{fig:kcore}(b), we present the empirical complementary cumulative distribution (CCDF) of the $k$--core values for spreaders and non--spreaders, respectively. The ego--net users have the largest k--core values in general, a strong evidence that they were the most influential part of the population. Moreover, Figure~\ref{fig:kcore_region} presents the k-core values of the nodes in the largest connected component of the region of influence. Again, the ego--net users have in general larger k-core values than the trolls.

\subsection{Cascade trees}
We now turn our attention to the diffusion cascades of the troll--URLs. First, in Figure~\ref{fig:cascades}(a) we present a general result, the \textit{relative first--appearance} of trolls and the ego--net users. For each URL, we rank the user IDs in descending order based on the date that they post their first tweet which contain the URL in question. This list is in fact the history of the diffusion of a given URL -- a series of consecutive instances of sharing a given URL through tweeting. It is just the list of user IDs that shared the URL, in chronological order. Recall that in our data, any URL might have been shared multiple times by the same user. The size of a given URL--diffusion is simply the number of user IDs belonging in the list (including the repetitions of IDs). In other words, the \textit{relative first--appearance} of a user in the URL--diffusion is the index of the user in the list divided by the length of this list. In Figure~\ref{fig:cascades}(a), we show the empirical cumulative distribution (CDF) for the \textit{relative position} of trolls and ego--net spreaders. Clearly, the largest part of the ego--net users appear before the trolls in the \textit{diffusion--lists}. Almost $85\%$ of the ego--net users appear first in the URL--diffusions.    

Next, we examine the URLs that have been shared by more than 100 distinct users. This selection led to 5,092 URL--diffusions. Then, based on the
method we described in the previous section, we construct the cascade trees for each URL--diffusion. In this way, we obtain 88,714
cascade trees for 5,084 URLs that have at least one non--empty cascade tree. In summary: (i) 4,125 cascades have at least one troll user
and all of them have at least one ego--net user; (ii) 64,525 cascades have at least one ego--net user and zero troll users. In the
cascade trees each user appears only once, hence the size of a cascade is equal to the number of distinct IDs belonging in the tree. In
Figure~\ref{fig:cascades}(b), we present the empirical complementary cumulative distribution (CCDF) of the \textit{relative
	first--appearance} of trolls and the ego--net users in the cascade trees. We observe that $80\%$ of the ego--net users appear very early
in the cascades.

The most viral cascades are those with both trolls and ego--net users (see Figure \ref{fig:cascades}(c)). At the same time, only 647 cascades had been initiated by troll users versus 54,111 by ego--net users. Roughly, $10\%$ of the cascades that had been initiated by trolls have a larger size, yet the truly viral cascades had been initiated mainly by the ego--net users (Figure~\ref{fig:cascades}(d)). This evidence that although the trolls participated in the viral cascades, they did not have a leading role in them. Instead, the primary source of the viral cascades was the real users.

\subsubsection{Influence}
Did the trolls have an influential role in the diffusion cascades? We address this question by computing for each user $i$ the \textit{influence--degree}, namely the number of real users that have been influenced by $i$ in the cascade trees he participates. Only 31 trolls have non--zero influence--degree. From them, 6 trolls have influence--degrees 978, 998, 2,961, 6,636, 9,040, and 46,224, respectively (Figure~\ref{fig:influence}(a)). Hence, only four trolls where truly influential. Regarding the ego--net users: 37,535 have
non--zero influence; 3,453 users, 511 users and 34 users have influence--degree larger than $10^{2}$, $10^{3}$ and $10^{4}$, respectively (Figure~\ref{fig:influence}(b)).

How similar with the troll accounts were the ego--net users? In order to estimate this, we use the classifications scores of the troll--category for the ego--net users with $\text{influence--degree} > 100$. We have 3,453 ego--net users with $\text{influence--degree}>100$, where 21 users were in the training set and 3,432 in the test set. Hence, we present the classification
scores for the 3,432 users, only. The $80\%$ of the users have classification score smaller than 0.6 (see Figure~\ref{fig:influence}(c)). Moreover, in Figure~\ref{fig:influence}(d) we show the scatter plot of the classification scores and 
the influence-degree for the ego--net real users with influence larger than 100. The classification scores -- the similarity of the
real users with the trolls in terms of tweets content -- represent the independent variable while the influence--degree the dependent one. The most influential users have low classification scores. Furthermore, the Pearson and Spearman correlation coefficients are ($r= -0.031$, $\text{p-value} = 0.067$) and ($r_{s}=0.021$, $\text{p-value} = 0.21$), respectively. Hence, there is neither a linear nor a monotonic relationship between the two variables.   

\subsubsection{Structural virality}
We conclude the analysis with the computation of the \textit{structural virality} for each cascade tree. The structural virality of a
cascade tree $T$ with $n>1$ nodes is the average distance between all pairs of nodes in a cascade \cite{Goel2015:virality}. That is:

\begin{equation}\label{eq:virality}
\nu(T) = \frac{1}{n(n-1)}\sum_{i=1}^{n}\sum_{j=1}^{n}d_{ij}
\end{equation}

\noindent where $T$ is the cascade tree with $n$ nodes and $d_{ij}$ is the shortest path between the nodes $i$ and $j$.
The intuition is that $\nu(T)$ is the average depth of nodes when we consider all nodes as the root of the cascade. 

We compute the structural virality for the 88,714 cascade trees of 5,084 URLs that were discussed previously. We examined two cases: (i) the original cascades (88,714); (ii) the cascade trees that we have reconstructed for the case where the trolls have been removed from both the graph and the 5,084 URL--diffusions. The exclusion of the trolls results in 95,539 cascade trees. This indicates that a portion
of the original trees broke in smaller trees, and probably some trolls act as ``bridges''. That is, they connect paths of the overall diffusion flow. On the other hand, it seems that their role was not substantial. The exclusion of the trolls did not affect the distribution of the virality values (see Figure~\ref{fig:virality}). 

\subsection{Summary of the results}
For all the metrics we have applied, the trolls as a whole have an inferior role compared to the ego--net users. The ego--net users,
namely the real users that had at least one interaction with the trolls, were the most influential population and the source of the viral cascades, as well. Table~\ref{tab:topk} provides the specifics.

\begin{table}[hbt]
	\centering
	\caption{Top--k results: trolls vs ego--net \textit{spreaders}}
	\label{tab:topk} 
	\setlength\tabcolsep{6pt}
	\begin{tabular}{l | l | l}
		Metrics                              & trolls  &  ego--net       \\\hline
		
		Popularity:	$\text{in--degree} > 10^{3}$         &  12     &  5,223   \\\hline
		
		Sociability: $\text{out--degree} > 10^{3}$         &  3     &  21,887   \\\hline
		
		Nodes in the largest k-core        &    7     &  3,552   \\\hline
		Source node (``patient--zero''):          &        &           \\
		Number of cascades                   &   647  & 54,111    \\ \hline
		
		Source node: number of cascades                    &       &           \\ 
		with $\text{cascade size} > 10^{3}$   & 12   &  725  \\\hline
		
		$\text{influence--degree} > 10^{3}$   &   4       &    511  \\
		\hline		
	\end{tabular}	                                  
\end{table}

Furthermore, we have derived the following statistics:
\begin{enumerate}
	\item Only 12 trolls but 5,223 ego--net users have in--degree larger than $10^{3}$ (multigraph).
	\item Only 3 trolls but 21,887 ego--net users have out--degree larger than $10^{3}$ (multigraph).  
	\item Only 7 trolls but 3,552 ego--net users belong to the largest k-core region of the graph (max k-core value = 854).
	\item Only 647 cascades had been initiated by trolls but 54,111 by ego--net users.
	\item Only 12 viral cascades (size larger than $10^{3}$) had been initiated by trolls but 725 by ego--net users.
	\item Only 4 trolls had influence--degree larger than $10^{3}$ but 511 for the ego--net users.
	\item The most influential ego--net users have low similarity with the trolls.
\end{enumerate} 

Finally, although the trolls participated in viral cascades, their role was not substantial. Removing the trolls from the graph as well as from the URL--diffusions did not affect the distribution of the structural virality values. 

\section{Conclusion}\label{sec:conclusions} 
In this paper, we have extensively studied the influence that state--sponsored troll Twitter accounts had during the 2016 US presidential election. We analyzed a very large graph which represents the interactions between trolls and real users and we concentrated our attention to the region of influence, namely the well-connected part of the graph where trolls could have influenced real users. We present strong evidence that the trolls' activity was not the source of the viral cascades. The authentic users who had close proximity with the trolls were the most influential part of the population and their activity was the driving force of the viral cascades.

\section{ Acknowledgments}
We are grateful to Twitter for providing access to the unhansed version of the trolls' ground truth dataset. This project has received funding from the European Union's Horizon 2020 Research and Innovation program under the Cybersecurity CONCORDIA project (Grant Agreement No. 830927).

\section{Appendix}\label{sec:appendix} 
Table \ref{tab:track_terms} presents the 77 track terms used in Twitter API. 

\begin{table}[hbt]
	\centering
	\caption{The track terms used in Twitter API}\label{tab:track_terms}
	\setlength\tabcolsep{2pt}
	\begin{tabular}{l | l }
		ben\%20carson &
		bencarson  \\
		bernie\%20sanders & 
		bernie2016  \\
		bettercandidatethanhillary &
		carlyfiorina2016 \\
		carson2016 &
		clinton \\
		clinton2016 &
		cruz2016 \\
		cruzcrew &
		cruzintocaucus \\
		donaldtrump &
		donaldtrump2016 \\
		dumptrump &
		election2016 \\
		feelthebern &
		fiorina \\
		fiorina2016 &
		fitn \\
		heswithher &
		hilaryclinton \\
		hillary2016 &
		hillaryclinton \\
		hrc &
		huckabee \\
		huckabee2016 &
		imwithher \\
		iwearebernie &
		jill\%20stein \\
		jillstein &
		johnkasich \\
		kasich &
		kasich2016 \\
		kasich4us &
		letsmakeamericagreatagain \\
		makeamericagreatagain &
		makeamericawhiteagain \\
		marco\%20rubio &
		marcorubio \\
		martinomally &
		mikehuckabee \\
		nevertrump &
		newyorkvalues \\
		nhpolitics &
		nhpsc \\
		omalley &
		paul2016 \\
		primaryday &
		randpaul2016 \\
		readldonaldtrump &
		realdonaldtrump \\
		redstate &
		rick\%20santorum \\
		ricksantorum &
		rubio2016 \\
		rubiowa &
		sensanders \\
		sentedcruz &
		stein2016 \\
		teamKasich &
		teamcarly \\
		teamclinton &
		teamcruz \\
		teamhillary &
		teammarco \\
		teamrubio &
		teamtrump \\
		ted\%20cruz &
		tedcruz \\
		the\%20donald &
		thedonald \\
		therealdonaldtrump &
		trump \\
		trump2016 &
		trumptrain \\
		unitedblue & \\ 
		\hline		
	\end{tabular}	                                  
\end{table}

\bibliography{Arxiv_Salamanos2019v2}
\bibliographystyle{abbrv}

\end{document}